\documentclass[12pt]{article}
\usepackage{axodraw}
\usepackage{amsmath}
\usepackage{amssymb}
\usepackage{cite}
\usepackage[]{hyperref}
\usepackage{tikz-cd}
\input xy
\xyoption{all}

\newcommand{\M}{{\overline{\cal M}}}

\numberwithin{equation}{section}

\begin{document}

\vspace*{0.5in}

\begin{center}

{\large\bf Global aspects of moduli spaces of 2d SCFTs}

\vspace{0.25in}

Ron Donagi$^1$, Mark Macerato$^1$, Eric Sharpe$^2$ \\

\vspace*{0.2in}

\begin{tabular}{cc}
{\begin{tabular}{l}
$^1$ Department of Mathematics\\
University of Pennsylvania\\
209 South 33rd Street\\
Philadelphia, PA  19104-6395
\end{tabular}
} &
{\begin{tabular}{l}
$^2$ Department of Physics MC 0435 \\
850 West Campus Drive \\
Virginia Tech \\
Blacksburg, VA  24061
\end{tabular}
}
\end{tabular}

{\tt donagi@math.upenn.edu},
{\tt macerato@sas.upenn.edu},
{\tt ersharpe@vt.edu}\\

$\,$

\end{center}

The Bagger-Witten line bundle is a line bundle 
over moduli spaces of two-dimensional SCFTs, related to the Hodge line
bundle of holomorphic top-forms on Calabi-Yau manifolds.  
It has recently been a subject
of a number of conjectures, but concrete examples have proven elusive.
In this paper we collect several results on this structure, including
a proposal for an
intrisic geometric definition over moduli spaces of Calabi-Yau manifolds
and some additional concrete examples.  
We also conjecture a new criterion for UV completion of
four-dimensional supergravity theories in terms of properties of the
Bagger-Witten line bundle.

\begin{flushleft}
June 2019
\end{flushleft}

\newpage

\tableofcontents

\newpage

\section{Introduction}

The Bagger-Witten line bundle over a moduli space of SCFTs
was originally introduced in
\cite{bw} to understand a subtlety in four-dimensional supergravity
theories, namely how the theory is well-defined across coordinate patches
on the target (moduli) space.  The resolution of this issue involved observing
that the fermions of the supergravity theory couple to a line bundle,
the Bagger-Witten line bundle, over the moduli space of scalar vevs.
Since then, analogues of the Bagger-Witten line bundle have also been
observed in superconformal field theory constructions, see 
{\it e.g.} \cite{ps,distsigma}, and \cite{Bershadsky:1993cx} where
it arises as a line bundle of Fock vacua over the moduli space.

Despite these general observations, not that much is known about
details of the Bagger-Witten line bundle.  For example, no explicit
examples were known until relatively recently \cite{mebw,ds17},
when the Bagger-Witten line bundle was computed for moduli spaces of
elliptic curves \cite{mebw}, a computation that involved slightly
extending the moduli space, and the closely related Hodge line bundle
was computed for moduli spaces of certain toroidal orbifolds \cite{ds17}. 
Other recent work (see {\it e.g.} \cite{Gomis:2015yaa} and references
therein) has argued that the Bagger-Witten line bundle is flat,
consistent with the known examples, in which the Bagger-Witten line bundle
is flat but nontrivial.

One reason to try to understand such structures and moduli spaces
in general is that they determine and are determined by
dualities.
For example, moduli spaces
of elliptic curves often arise in various contexts in string theory,
realized by some fields related to a $\tau$ parameter.  The $\tau$
parameter itself lives on the upper half plane, but duality symmetries
exchange values of $\tau$ related by 
$PSL(2,{\mathbb Z})=SL(2,{\mathbb Z})/Z$, 
where $Z \cong {\mathbb Z}_2$ is the center, and so because of those
duality symmetries, we identify $\tau$ with a parameter on the moduli
space of elliptic curves.  This illustrates how
knowing the moduli space is equivalent to
knowing the (self-)dualities.
In type IIB string theories in ten dimensions, for example,
such a $\tau$ encodes the complexified dilaton.  While the center $Z$ of 
$SL(2,{\mathbb Z})$  acts trivially on $\tau$, its action on other fields of the theory turns out to be non-trivial. 
For this reason the physically meaningful duality group on bosons is
 $SL(2,{\mathbb Z})$ rather than just $PSL(2,{\mathbb Z})$.   
Further, to describe the action of the duality group on
fermions \cite{meta}, 
one must extend $SL(2,{\mathbb Z})$ to the metaplectic group $Mp(2,{\mathbb Z})$,
reflecting the fact that it is a duality, and the quotient
\begin{equation}
\left[ \left(\mbox{upper half plane}\right) / Mp(2,{\mathbb Z})
\right]
\end{equation}
is the stack over which the Bagger-Witten line bundle on a moduli space
of elliptic curves is defined \cite{mebw}.

The purpose of this paper is to propose an intrinsic definition of the Bagger-Witten line bundle 
and of the moduli space on which it lives, 
and to explore some further aspects of both, including computing
more examples, which are in short supply.
The correct moduli ``space" turns out to be a ${\mathbb Z}_2$-gerbe 
over the usual moduli space, which physically implies subtle
${\mathbb Z}_2$ self-duality actions on fermions of the low-energy theories.

We begin in section~\ref{sect:review} by reviewing basic aspects of the
Bagger-Witten line bundle, and its structure in the special case of
 a (${\mathbb Z}_2$ gerbe
over a) moduli space
of elliptic curves, which forms a prototype for much of our discussion.

In section~\ref{sect:defn}, we propose our intrinsic definition of the
Bagger-Witten line bundle.  The Bagger-Witten
line bundle was originally defined in terms of transformations of
K\"ahler potentials across moduli spaces, but this is not an intrinsic
definition.  We propose an intrinsic definition as a bundle of
covariantly-constant spinors, in close analogy with the Hodge line bundle
of holomorphic top-forms.  We give several mathematically-rigorous definitions,
show they are equivalent, and also check that for moduli spaces of
elliptic curves, they specialize to the Bagger-Witten line bundle
given in \cite{mebw}.  

Known examples of Bagger-Witten line bundles are in very short supply,
so next we turn to the construction of more explicit examples.
In section~\ref{sect:warmupmodels} we warm up by describing the Bagger-Witten
line bundle over some models of moduli spaces of
Calabi-Yau manifolds that are comparatively easy to describe.
In section~\ref{sect:toroidalexs} we construct
the Bagger-Witten line bundles over the moduli spaces of
toroidal orbifolds discussed in \cite{ds17}.

In section~\ref{sect:weakgrav}, we outline some mathematics conjectures
regarding Bagger-Witten line bundles.  In particular, we observe that
in all currently known examples, the Bagger-Witten and Hodge line
bundles over moduli spaces of Calabi-Yau manifolds are flat but
nontrivial, and suggest that this may be a criterion for determining
whether a given supergravity theory has a UV completion.
In sections~\ref{sect:superpotential:27:3} and \ref{sect:fi} we 
clarify some puzzles regarding applications of these
ideas to spacetime superpotentials and Fayet-Iliopoulos parameters.
Finally, in appendix~\ref{app:picard}, we collect a number of 
technical properties of moduli spaces of elliptic curves with level 
structures, which are used in section~\ref{sect:warmupmodels}.

Related notions, motivated by higher-dimensional conformal field
theories, have been recently discussed in 
\cite{Donagi:2017vwh,Tachikawa:2017aux}.

In passing, we should briefly observe some relations to string
dualities.  
One of our arguments is that typical moduli spaces of Calabi-Yau's
pertinent for physics (Calabi-Yau's with spin structure) 
have a ${\mathbb Z}_2$ gerbe structure, corresponding to an extra
local ${\mathbb Z}_2$ quotient.  Now, moduli spaces are tightly
interlinked with self-dualities:  subtleties in a moduli space
should correspond physically to subtleties in self-dualities of
the theories in question.  Here, it is worth observing that there
are closely related ${\mathbb Z}_2$'s which appear in duality
group actions on fermions.  One of the authors observed this
in \cite{meta} for the case of moduli spaces of elliptic curves,
in their various manifestations in string theory, and in fact
there are closely related subtleties generically in supergravity
theories.  Generally, a supergravity theory has a global symmetry
$G$ and a local symmetry $H$, which is a ${\mathbb Z}_2$ extension
of a local symmetry $H'$ which acts on bosons.  (The extra
${\mathbb Z}_2$ acts on the fermions of the theory.)  For example,
in four-dimensional $N=8$ supergravity, the global symmetry group
is $E_{7(7)}$, with maximal compact subgroup $SU(8)/{\mathbb Z}_2$.
The bosonic degrees of freedom form representations of
$E_{7(7)}$, and do not see the ${\mathbb Z}_2$,
while the fermions are in representations of
$SU(8)$ which do see the ${\mathbb Z}_2$ subgroup.
For more information, see for example
\cite{Hull:2007zu,Keurentjes:2003hc}.

\section{Review of Bagger-Witten and moduli spaces of elliptic curves}
\label{sect:review}

Bagger-Witten line bundles were originally discovered in four-dimensional
$N=1$ supergravity theories \cite{bw}.  Briefly, it was observed there that
in such supergravities,
K\"ahler potential transformations
\begin{equation}
K \: \mapsto \: K + f + \overline{f}
\end{equation}
are not a symmetry of the theory, unlike in rigid supersymmetry, unless
they are combined with chiral\footnote{
Since these are chiral rotations, there are potential anomalies
(see {\it e.g.} \cite{ds,Tateishi:2018rnq}),
but here we shall focus on the classical analysis.
} rotations of the gravitino $\psi_{\mu}$
and scalar superpartners $\chi^i$ (see {\it e.g.} \cite{wb}[(23.9)]):
\begin{equation}
\psi_{\mu} \: \mapsto \: \exp\left( - \frac{i}{2} {\rm Im}\, f \right)
\psi_{\mu}, \: \: \:
\chi^i \: \mapsto \: \exp\left( + \frac{i}{2} {\rm Im}\, f \right) \chi^i.
\end{equation}
If the theory has no gauge symmetries, then as argued in \cite{bw}
the phase factors above close on triple overlaps and so
define a
line bundle over the moduli space of scalars in the supergravity theory,
known as the Bagger-Witten line bundle.
If the four-dimensional theory does have gauge symmetries, then on triple
overlaps the transition functions need not close, and so instead of
an honest line bundle, one has a line bundle on a gerbe,
what is sometimes deemed a
`fractional' line bundle (see \cite{mebw,hs} for a discussion
of this issue, and \cite{ajmos} for more information on fractional
line bundles).  Such objects can be understood as line bundles on
stacks, and so we will treat them uniformly as line bundles in this paper.

More formally, we can write the above as\footnote{
In the expressions above, we use the fact that as $C^{\infty}$
bundles, $\overline{\cal L} \cong {\cal L}^{-1}$.
}
\begin{eqnarray}
(\psi_{\mu}) & \in & \Gamma\left( TX \otimes
\phi^* {\cal L}_{\rm BW}^{+1/2} \otimes \phi^* \overline{\cal L}_{\rm BW}^{
-1/2} \right) \: \cong \:
\Gamma\left( TX \otimes
\phi^* {\cal L}_{\rm BW} \right), \\
(\chi^i) & \in & \Gamma\left( T M \otimes
\phi^* {\cal L}_{\rm BW}^{-1/2} \otimes
\phi^* \overline{\cal L}_{\rm BW}^{+1/2} \right)
\: \cong \:
 \Gamma\left( T M \otimes
\phi^* {\cal L}_{\rm BW}^{-1}
\right),
\end{eqnarray}
where $X$ is the four-dimensional spacetime, $\phi: X \rightarrow M$
the scalar vevs, and the isomorphisms above are isomorphisms of
$C^{\infty}$ bundles. 

In any event, the definition of the Bagger-Witten line bundle above --
in terms of transformation properties of the K\"ahler potential -- is
not particularly intrinsic.  We shall attempt to remedy this matter,
by proposing an intrinsic definition in section~\ref{sect:defn}.

Although the Bagger-Witten line bundle was originally discovered in
four-dimensional $N=1$ supergravity, an analogue exists over moduli spaces
of two-dimensional $N=2$ SCFTs \cite{ps,distsigma}, corresponding to
a (fractional) line bundle of $U(1)_R$ rotations.  In general, any global
symmetry in a worldsheet theory will define some sort of bundle
over the moduli space of SCFTs, with transition functions defined by
the symmetry group actions across patches on the moduli space.  In the
present case, the Bagger-Witten bundle corresponds to the two-dimensional
$U(1)_R$ rotations of the $N=2$ SCFT.

In that vein, a Bagger-Witten line bundle can be understood
as a bundle of chiral Ramond Fock vacua in the worldsheet theory.
(For example, the line bundle of Fock vacua discussed in
\cite{Bershadsky:1993cx} is a Bagger-Witten line bundle.)

As explained in {\it e.g.} \cite{ps} and \cite{mebw}[section 2.2], 
another way of thinking about the Bagger-Witten line bundle is
as the bundle of spectral flow operators 
${\cal U}_{1/2}$ over the SCFT moduli space, where ${\cal U}_{1/2}$
rotates from R to NS and vice-versa.  In a string compactification to
four dimensions, this operator
is the essential piece of the vertex operator for the spacetime
supercharge, which is how this worldsheet description of Bagger-Witten
is connected to the target 
spacetime interpretation in terms of phase rotations of
the gravitino, scalar superpartners, and so forth.  It is also one of the
reasons why we later propose that the Bagger-Witten line bundle be
understood geometrically as a line bundle of covariantly constant
spinors, as we shall describe in section~\ref{sect:defn}.
  
Spectral flow by $\theta=1$
(in the conventions of \cite{Lerche:1989uy}) is
described by the operator ${\cal U}_{1}$, which corresponds in a 
Calabi-Yau compactification to the holomorphic top-form.
Briefly, the upshot is that the bundle of spectral flow operators
${\cal U}_1$ corresponds to the bundle of holomorphic top-forms,
namely the Hodge line bundle ${\cal L}_H$, which is therefore
related to the Bagger-Witten line bundle ${\cal L}_{\rm BW}$ as
${\cal L}_{\rm BW}^{\otimes 2} \cong {\cal L}_H$.

In any event, the descriptions of the Bagger-Witten line bundle above --
in terms of transformation properties of the K\"ahler potential, or
as a line bundle of Fock vacua -- are not particularly amenable to
an intrinsic mathematical understanding.
We shall attempt to remedy this matter,
by proposing an intrinsic definition more nearly suitable for
mathematicians in section~\ref{sect:defn}.

So far we have outlined generalities.  Next, we will
briefly review the highlights of the structure in the special case
of moduli
spaces of elliptic curves, which provide explicit examples of
these ideas.  The moduli space over which the
Bagger-Witten line bundle is well-defined is a slight variation of the
usual moduli space, so we will first carefully review the
usual moduli space.

First, as is well-known, $SL(2,{\mathbb Z})$ acts on the complex
structure parameter $\tau$ of an elliptic curve as
\begin{equation}
\tau \: \mapsto \: \frac{a \tau + b}{c \tau + d}
\end{equation}
for
\begin{equation}
\left( \begin{array}{cc}
a & b \\
c & d \end{array} \right) \: \in \: SL(2,{\mathbb Z}),
\end{equation}
a map which is invariant under the center
$\{\pm 1\} \subset SL(2,{\mathbb Z})$.
For this reason, we ordinarily think of the moduli space of
(complex structures on) elliptic curves as the quotient
\begin{equation}
{\cal M}^{\rm red} \: \equiv \:
\left( \mbox{upper half plane} \right) / PSL(2,{\mathbb Z}),
\end{equation}
where $PSL(2,{\mathbb Z}) = SL(2,{\mathbb Z})/\{\pm 1\}$.

Now, a holomorphic coordinate $z$ on an elliptic curve also
transforms under $SL(2,{\mathbb Z})$ at the same time as $\tau$,
to ensure that the relation $z \cong z + m \tau + n$ is preserved,
for $m, n \in {\mathbb Z}$.  The pair transform as \cite{hainrev}[section 2.3]
\begin{equation}
(z, \tau) \: \mapsto \: \left( \frac{z}{c \tau + d},
\frac{a \tau + b}{c \tau + d} \right).
\end{equation}

The holomorphic top-form on an elliptic curve is simply $dz$, so we see
that under $SL(2,{\mathbb Z})$,
\begin{equation}
dz \: \mapsto \: \frac{dz}{c \tau + d}.
\end{equation}
In particular, under the center $\{ \pm 1 \} \subset SL(2,{\mathbb Z})$,
$dz$ is not invariant but rather $dz \mapsto - dz$.  As a result,
the bundle of holomorphic top-forms does not descend to the moduli space
${\cal M}^{\rm red}$ above, as it is a $PSL(2,{\mathbb Z}) = SL(2,{\mathbb Z})
/ \{ \pm 1 \}$ quotient and $dz$ is not invariant under $\{ \pm 1 \}$.
Instead, to have a moduli space over which the line bundle of holomorphic
top-forms is defined, we must instead work on the (stack) quotient
\begin{equation}
{\cal M} \: \equiv \:
\left[ \left( \mbox{upper half plane} \right) / SL(2,{\mathbb Z})
\right].
\end{equation}
Although $PSL(2,{\mathbb Z})$ and $SL(2,{\mathbb Z})$ define the same quotient
space, since the ${\mathbb Z}_2$ center acts trivially, stacks keep
track of even trivially-acting group quotients, and so the $PSL(2,{\mathbb Z})$
and $SL(2,{\mathbb Z})$ quotient stacks are different from one another.
(In passing, a moduli space of elliptic curves with fixed holomorphic
top-form, not just a well-defined line bundle of holomorphic top-forms,
would be the total space of the Hodge line bundle over the moduli space
above, minus the zero section.)

Now, suppose we are describing a moduli space of (2,2) supersymmetric
nonlinear sigma models with target an elliptic curve with trivial
spin structure.  (The spin structure of the target space enters
into the Fock vacuum, as we shall discuss later.)
If we want a moduli space over which the bundle of chiral R sector
vacua is well-defined, we must work harder still.  Since we have a single
complex fermion $\psi$, in a chiral R sector there are two Fock
vacua, typically denoted $| \pm \rangle$, and defined by
\begin{equation}
\psi | + \rangle \: = \: 0, \: \: \:
\psi |- \rangle \: = \: | + \rangle, \: \: \:
\overline{\psi} | - \rangle \: = \: 0.
\end{equation}
Following standard methods for orbifolds,
under $\psi \mapsto - \psi$, these vacua transform as
\begin{equation}
| \pm \rangle \: \mapsto \: \pm \exp(i \pi/2) | \pm \rangle,
\end{equation}
which is consistent with the relation between the vacua defined by
multiplication by $\psi$ above.  However, because of those transformation
laws, note that a line bundle of Fock vacua is not well-defined on
${\cal M}$, the $SL(2,{\mathbb Z})$ quotient, as the action of the center
$\{ \pm 1 \}$ on the vacua generates a phase that does not square to one
and so is not consistent with the group law.  In order to get a moduli space
over which the Fock vacua are well-defined globally, we must take a further
quotient.

To understand the further quotient, let us now repeat this discussion in
greater generality.  In a nonlinear sigma model on a target space $X$,
the chiral R sector vacua in the RNS formalism produce spacetime spinor
indices.  Now, on a K\"ahler manifold, the spinor bundle can be described
as\footnote{
Note that the wedge power is interpreted as the complex exterior power of
$TX$, not the real exterior power.
}  \cite{lm}[equ'n (D.16)] 
\begin{equation}
\wedge^* TX \otimes \sqrt{K_X},
\end{equation}
which is represented by the Fock vacua:  the $\wedge^* TX$ corresponds to
multiplying by various $\psi$'s to interchange the Fock vacua, and
a given Fock vacuum (in these conventions, corresponding to the one 
annihilated by
$\overline{\psi}$'s) couples to $\sqrt{K_X}$, and so, for example,
encodes the target-space spin structure.
(See also {\it e.g.} \cite{Sharpe:2013bwa}.)

In the present case, for $X$ an elliptic curve with trivial spin
structure,
the fact that the Fock vacuum couples to $\sqrt{K_X}$ means that
under an $SL(2,{\mathbb Z})$ transformation, since
\begin{equation}
dz \: \mapsto \: \frac{dz}{c \tau + d},
\end{equation}
a given Fock vacuum transforms as $\sqrt{dz}$, {\it i.e.}
\begin{equation}
\sqrt{dz} \: \mapsto \: \pm \frac{ \sqrt{dz} }{\sqrt{ c \tau + d} }.
\end{equation}
This is precisely the action of the metaplectic group $Mp(2,{\mathbb Z})$,
the unique nontrivial central extension of $SL(2,{\mathbb Z})$ by
${\mathbb Z}_2$, and so we see that to define a line bundle of chiral Ramond
Fock vacua (the Bagger-Witten line bundle)
over the moduli space of elliptic curves, we must define the moduli
space with an $Mp(2,{\mathbb Z})$ quotient:
\begin{equation}
{\cal M}^S \: \equiv \:
\left[ \left( \mbox{upper half plane} \right) / Mp(2,{\mathbb Z}) \right].
\end{equation}
This matter was explored in \cite{mebw}, which proposed that the
$Mp(2,{\mathbb Z})$ quotient above
is the more-nearly correct moduli space of complex
structures for SCFTs for elliptic curve targets.  Of course, this also
implies a modification of T-duality, which was propagated through other
string dualities in \cite{meta}.

For completeness, we should add that for nontrivial spin structures on the
target elliptic curve, matters are more complicated.
For example, although $Mp(2,{\mathbb Z})$ maps the trivial spin structure
to itself, it will in general
exchange spin structures on the target-space elliptic curve.  Furthermore,
precisely one of those target-space spin structures will be compatible
with low-energy supersymmetry:  only in the target-space RR sector will
there exist a globally-defined nowhere-zero covariantly constant spinor
that squares to the holomorphic top-form.  (For other target-space spin
structures, the spinor bundle will not have a nowhere-zero covariantly
constant section.)
To construct a moduli space of elliptic curves on which the trivial
spin structure naturally enters, one considers an $Mp(2,{\mathbb Z})$ 
quotient; however, for an analogous moduli space of elliptic curves with
nontrivial spin structures, the correct moduli space would be more
complicated.

In passing, the reader might be curious about the multiplicity of
these bundles, due to the fact that the moduli spaces are not simply-connected.
For example, the moduli stack ${\cal M}$ has fundamental group
$SL(2,{\mathbb Z})$, and the moduli stack
${\cal M}^S$ has fundamental group $Mp(2,{\mathbb Z})$.  When a K\"ahler
manifold is not simply connected, one has a multiplicity of square roots
counted by homomorphisms $\pi_1 \rightarrow {\mathbb Z}_2$, and the
Bagger-Witten line bundle ${\cal L}_{\rm BW}$ is a square root of
the Hodge line bundle ${\cal L}_H$.  However, although the moduli stacks
are not simply-connected, there are no nontrivial homomorphisms
from
either $SL(2,{\mathbb Z})$ or $Mp(2,{\mathbb Z})$ to ${\mathbb Z}_2$,
hence for example there is no ambiguity in the definition of
${\cal L}_{\rm BW}$ as a square root of ${\cal L}_H$ in this case.

Finally, we should make one additional comment regarding what constitutes
a moduli spaces of SCFTs.  In the discussion above, we focused on finding
a moduli space over which the line bundle of Fock vacua is well-defined.
However, depending upon the circumstances, one may wish to impose
additional constraints.  Suppose one wishes to build a moduli space over
which physical fields of nonlinear sigma models on Calabi-Yau targets are
well-defined.  Defining the worldsheet fermions globally over the moduli
space would further constrain it.  To see this, note that if the
Calabi-Yau has complex dimension $n$, then the holomorphic top-forms
have $U(1)_R$ charge $n$ when a single worldsheet fermion has 
$U(1)_R$ charge $1$.  Since the holomorphic top-forms couple to the Hodge
line bundle over the moduli space, the individual worldsheet fermions couple
to an $n$th root of the Hodge line bundle, whose defintion could in general
require taking a ${\mathbb Z}_n$ gerbe over the original moduli space.
For example, for Calabi-Yau three-folds, the line bundle of Fock vacua
require a ${\mathbb Z}_2$ gerbe over the original moduli space,
and the worldsheet fermions require a ${\mathbb Z}_3$ gerbe over the moduli
space.  Altogether, one would appear to need a ${\mathbb Z}_6$ over the
moduli space in order to define both Fock vacua and worldsheet fermions 
globally.  (Happily, for moduli spaces of elliptic curves, matters are
simpler, as the worldsheet fermions and holomorphic top-form have the
same $U(1)_R$ charge and so both couple to the Hodge line bundle.)
In this paper, we have decided to focus just on building
moduli spaces of abstract SCFTs pertinent for Bagger-Witten line bundles, 
for which we need to make sense of
Fock vacua, but not the worldsheet fields in any given Lagrangian realization
of a family of SCFTs.

\section{Proposal for an intrinsic definition}
\label{sect:defn}

\subsection{Outline and intuition} \label{intuition}

The Bagger-Witten line bundle was originally defined in
\cite{bw} in terms of the transition functions of the
K\"ahler potential over the moduli space.
In this section we will propose a more intrinsic definition,
as a line bundle of covariantly constant spinors,
just as the Hodge line bundle is a line bundle of holomorphic top-forms
on a moduli space of Calabi-Yau manifolds.
To be clear, a bundle of spinors should typically
have rank greater than one, but 
 the essential part of
the variation -- the variation of a single, covariantly constant, spinor --
is determined by a line bundle.  In this subsection, we will
further outline this description, at an intuitive level.
In the next subsection, we will
give a formal technical definition suitable for mathematicians.  
In the final subsection,
we will resolve a puzzle arising when comparing our definition
to that in \cite{bw}.

Now, as outlined, this definition of the Bagger-Witten line bundle
is ambiguous, as on a Calabi-Yau $n$-fold
of maximal holonomy\footnote{
By `maximal' holonomy we mean $SU(n)$ holonomy on a Calabi-Yau $n$-fold.
For example, as Calabi-Yau threefolds, $T^6$ and $K3 \times T^2$ have
holonomy that is a proper subgroup of $SU(3)$, hence enhanced supersymmetry
and additional covariantly constant spinors.
}, there are always two covariantly constant spinors 
(or one and its conjugate, see for example \cite{Gauntlett:2003cy}).
As a result, there should always be at least two line bundles which could be
labelled as a line bundle of covariantly-constant spinors, and more
than that if the holonomy is submaximal.
Indeed, we will argue that there are
multiple Bagger-Witten line bundles, reflecting the multiple
possible covariantly-constant spinors, which will tie into recent
discussions of flatness of Bagger-Witten line bundles.
In simple examples, we will argue that
one spinor couples to
${\cal L}_{\rm BW}$, the other to ${\cal L}_{\rm BW}^{-1}$, related by
${\cal L}_{\rm BW}^{-1} \otimes {\cal L}_H \cong {\cal L}_{\rm BW}$,
reflecting the statement that ${\cal L}_{\rm BW}^{\otimes 2} \cong
{\cal L}_H$.  As we could replace the Hodge line bundle with its
dual, the line bundle of top polyvector fields, we see that both 
Bagger-Witten line bundles ${\cal L}_{\rm BW}^{\pm}$ can be interpreted
as square roots of some Hodge line bundle.

All that said, although there are multiple line bundles that could
reasonably be labelled a ``line bundle of covariantly-constant spinors,''
we will give a canonical line bundle in the next subsection.
In the rest of this paper, when we speak of ``the'' Bagger-Witten
line bundle, we are referring to that canonical choice.

Furthermore, we should also point out that not all possible square roots
of the Hodge line bundle may correspond to line bundles of
covariantly-constant spinors.  For example, consider a moduli space
of elliptic curves, as reviewed in \cite{hainrev}.
The Picard group of the stack is ${\mathbb Z}_{12}$, generated by the
Hodge line bundle.  The stack over which the Bagger-Witten line bundle
is defined \cite{mebw} is a ${\mathbb Z}_2$ gerbe on the moduli stack
of elliptic curves, with Picard group ${\mathbb Z}_{24}$.
If we let $g$ denote a generator of the Picard group of this
gerbe, then the pullback of the Hodge line bundle and its inverse
are $g^2$ and $g^{22}$, which have four square roots:
$g$, $g^{11}$, $g^{13}$, and $g^{23}$.  (This is because, for example,
$2 * 13 = 26 \equiv 2 \: {\rm mod} \: 24$.)  However, this is more
roots than we can obviously associate to covariantly-constant spinors;
it seems that
not every square root of the Hodge line bundle can be a Bagger-Witten
line bundle.

To begin to understand these matters, let us illustrate the proposed
definition in a little more detail.
Recall from \cite{gsw}[section 15.5] that spinors on Calabi-Yau's can
be
constructed
as follows.  Begin with a highest-weight state (in the sense of representation
theory) $| \Omega \rangle$, annihilated by gamma matrices $\gamma^i$.
Then, spinors are constructed as
\begin{equation}
| \Omega \rangle, \: \: \:
\gamma^{\overline{\imath}} | \Omega \rangle, \: \: \:
\gamma^{ \overline{\imath} \overline{\jmath} } | \Omega \rangle, \: \: \:
\cdots.
\end{equation}
In the description of \cite{gsw}[section 15.5], $| \Omega \rangle$
corresponds to
a covariantly constant spinor, and 
we propose that\footnote{
It appears that the same conclusion cannot be obtained from the construction
of holomorphic top-forms from covariantly constant spinors, as
$\omega_{ijk} \propto \eta^{\dag} \gamma_{ijk} \eta$,
for $\eta$ a covariantly constant
spinor and $\gamma_{ijk}$ a three-index antisymmetric
product of gamma matrices.  Instead, the
$\gamma_{ijk}$ couples to the line bundle of holomorphic top-forms, as we shall
see below, and the $\eta^{\dag}$ couples to the dual bundles of $\eta$,
cancelling out each other's dependencies.
} the covariantly constant spinor transforms over
the Calabi-Yau moduli space as a section of a Bagger-Witten line bundle.
(Similarly, the `filled' state at the opposite end of the representation
is identified with the other covariantly constant spinor, and transforms
over the moduli space as a section of another Bagger-Witten line bundle.)

It may be helpful to also think about the gamma matrices in this language.
Consider first a single elliptic curve in two dimensions.
If one identifies
\begin{equation}
\gamma^0 \sim \sigma^1 \: = \: \left[ \begin{array}{cc}
0 & 1 \\
1 & 0\end{array} \right], \: \: \:
\gamma^1 \sim \sigma^2 \: = \: \left[ \begin{array}{cc}
0 & -i \\ i & 0 \end{array} \right].
\end{equation}
then viewed locally as objects over the moduli space of elliptic curves,
\begin{equation}
\gamma_z \: = \: \left[ \begin{array}{cc}
0 & s \\
0 & 0 \end{array} \right], \: \: \:
\gamma_{\overline{z}} \: = \: \left[ \begin{array}{cc}
0 & 0 \\ s^{-1} & 0 \end{array} \right],
\end{equation}
where $s$ transforms locally as a section of ${\cal L}_H$.  Gamma matrices over
the Calabi-Yau's discussed here could be obtained from tensor products of
the gamma matrices for elliptic curve factors above.
Thus, we see that the gamma matrices will map sections of ${\cal L}_H^{-1}$
to ${\cal O}$ and conversely, consistent with the structure of the spin
operators over the moduli space as described above.

More generally, the holomorphic top form corresponds to the product
of gamma matrices that transform the highest-weight state
$| \Omega \rangle$ to the lowest-weight state -- so if the highest- and lowest-
weight states transform as a Bagger-Witten line bundle and its inverse,
then we have that ${\cal L}_{\rm BW} \cong {\cal L}_{\rm H} \otimes
{\cal L}_{\rm BW}^{-1}$, or ${\cal L}_{\rm BW}^{\otimes 2} \cong 
{\cal L}_{\rm H}$, as previously mentioned.

Let us consider the concrete example of $T^6$, with a complex structure
describing it as a product of three elliptic curves.
Now, $T^6$ has more than two covariantly constant spinors -- in
fact, it has $2^3 = 8$ covariantly constant spinors (or more accurately,
an eight-dimensional vector space of covariantly constant spinors), as we will
see explicitly next.

For simplicity, we will assume that the $T^6$ has a complex
structure such that it can be written as a product of three
elliptic curves, so that we can apply previous results.
In the worldsheet SCFT,
the spacetime spin operators
are given as
tensor products of the Fock vacua over the three elliptic curves,
{\it i.e.}
\begin{equation}
| \pm \rangle_1 \otimes | \pm \rangle_2 \otimes | \pm \rangle_3.
\end{equation}
Now, if the bundle of chiral Ramond
Fock vacua $| \pm \rangle$ over a single
moduli space $[ {\mathfrak h}/ \tilde{\Gamma}_1(2) ]$ is
${\cal L}_{BW}\oplus {\cal L}^{-1}_{BW}$,
then the vector bundle of spin operators over
the moduli space is
\begin{equation}
p_1^* \left( {\cal L}_{BW} \oplus {\cal L}^{-1}_{BW} \right) \otimes
p_2^* \left( {\cal L}_{BW} \oplus {\cal L}^{-1}_{BW} \right) \otimes
p_3^* \left( {\cal L}_{BW} \oplus {\cal L}^{-1}_{BW} \right),
\end{equation}
where the $p_i$ are the projections to the three factors.

A Bagger-Witten line bundle over the moduli space of this threefold
should simultaneously be two things:
first, a square root of the line bundle of holomorphic top-forms
\begin{equation}
p_1^* {\cal L}_H \otimes p_2^* {\cal L}_H \otimes p_3^* {\cal L}_H,
\end{equation}
and second, should appear as a factor determining the spin operator over
the moduli space.

Since the Bagger-Witten line bundle is a square root of the line
bundle of holomorphic top-forms, we can choose it to be
\begin{equation}
p_1^* {\cal L}_{BW} \otimes p_2^* {\cal L}_{BW} \otimes p_3^* {\cal L}_{BW}.
\end{equation}
We also see this structure emerge in the vector bundle of spin operators
by writing the vector bundle as
\begin{eqnarray*}
\lefteqn{
p_1^* \left( {\cal L}_{BW} \oplus {\cal L}^{-1}_{BW} \right) \otimes
p_2^* \left( {\cal L}_{BW} \oplus {\cal L}^{-1}_{BW} \right) \otimes
p_3^* \left( {\cal L}_{BW} \oplus {\cal L}^{-1}_{BW} \right)
} \\
& = & 
\left( p_1^* {\cal L}_{BW} \otimes p_2^* {\cal L}_{BW} \otimes p_3^* {\cal L}_{BW} \right) \otimes
\left(
p_1^*\left( {\cal O} \oplus {\cal L}_H^{-1} \right) \otimes
p_2^*\left( {\cal O} \oplus {\cal L}_H^{-1} \right) \otimes
p_3^*\left( {\cal O} \oplus {\cal L}_H^{-1} \right)
\right).
\end{eqnarray*}
Here, we have extracted the Bagger-Witten line bundle on the moduli space
of the threefold.  The fact that it appears as a factor in the spin operator
is consistent with statements that four-dimensional spinors -- the
gravitino, the scalar superpartners, the gaugino -- all couple to the
Bagger-Witten line bundle.

\subsection{Our definition}

So far we have given an intuitive outline of how one might
define the Bagger-Witten line bundle more intrinsically.
In this subsection we will give several equivalent mathematical definitions of the Bagger-Witten line bundle. 
We also check that in the case of elliptic curves it gives the correct object, 
described previously in terms of the metaplectic group.

\subsubsection{The stack of CYs with a root of the canonical bundle}  \label{root}

The moduli space over which the Bagger-Witten line bundle is defined
will be the moduli stack $\M$ of Calabi-Yau's  $X$ equipped with a ``square root of the canonical bundle."
By the latter we mean a trivializable line bundle $L$ together with an isomorphism $L^{\otimes 2} \cong K_X$. 
Given a Calabi-Yau $X$,
such a (trivializable) square root of $K_X$ is unique up to an isomorphism. 
But in general, this isomorphism is non-unique: 
in fact, there are two such isomorphisms commuting with the maps to $K_X$,
differing from each other by a sign change.
The moduli stack $\M$ of Calabi-Yau's  $X$ equipped with a square root of the Hodge bundle
is therefore a ${\mathbb Z}_2$-gerbe 
$p: \M \to {\cal M}$
over the  (ordinary) moduli stack  ${\cal M}$ of Calabi-Yau varieties.
Over ${\cal M}$ there is a well-defined Hodge line bundle ${\cal L}_H$
and a universal Calabi-Yau $\pi: {\cal U} \rightarrow {\cal M}$.
The pullback $\pi^* {\cal L}_H$ defines a line bundle over the universal family ${\cal U}$ 
whose restriction to any one Calabi-Yau fiber can be identified with the bundle of
holomorphic top-forms. 
(More generally, a line bundle can be identified with the pullback of its direct image if and only if it is trivializable.)
These objects on ${{{\cal M}}}$ can be pulled back to $\M$ via $p$: we have a Hodge bundle
${{\overline{{\cal L}_H}}} := p^*({\cal L}_H)$, and a universal family 
${\overline{\pi}} : {{\overline{{\cal U}}}} := p^* {\cal U} \to \M$ 
of Calabi-Yau's with a root of the Hodge bundle.
By definition, there is a line bundle ${\cal L}_{BW}$ on $\M$
whose pullback ${\overline{\pi}}^* {\cal L}_{BW}$ to ${{\overline{{\cal U}}}}$
encodes the corresponding root structures
over the family:
${\overline{\pi}}^* {\cal L}_{BW}$ is trivial along each fiber
of ${\overline{\pi}}$, and there is a natural isomorphism 
${\cal L}_{BW}^{\otimes 2} \cong {{\overline{{\cal L}_H}}}$.
This line bundle ${\cal L}_{BW}$ is our first version of the Bagger-Witten line bundle.



\noindent {\bf Remark} 
The structure we consider here is somewhat reminiscent of a spin structure, but differs from it in two significant ways: it is a square root of the canonical line bundle rather than of the tangent or cotangent vector bundles; and it is required to be trivializable.

\noindent {\bf Remark} 
Physically, as mentioned earlier, in a nonlinear sigma model on $X$,
the Fock vacuum couples to this line bundle.
The trivializability criterion ensures that for
non-simply-connected Calabi-Yau's, we are taking the spin structure
in which fermions are periodic around noncontractible loops,
corresponding to Ramond-sector boundary conditions.  This is the
sector in which the covariantly-constant spinor lives.

\subsubsection{A description by an atlas and relations}  \label{atlas}

An equivalent construction can be given as follows.
Let $U$ be an atlas on the moduli stack ${\cal M}$ of (ordinary)
Calabi-Yau varieties, with $R$ the relations defining ${\cal M}$
as a groupoid.  Let $\tilde{U}$ be the ${\mathbb C}^{\times}$
bundle on $U$, given by the punctured Hodge line bundle (corresponding
to nonzero holomorphic top-forms), and let $\tilde{R}$ denote the fiber product
$\tilde{R} := R \times_U \tilde{U} \times_U \tilde{U}$,
so that a point of $\tilde{R}$ is a point of $R$ plus a pair of
holomorphic top-forms.
This $\tilde{R}$ is a
${\mathbb C}^{\times} \times {\mathbb C}^{\times}$ bundle on $R$,
and the two projections of $R$ to $U$ lift to two projections of $\tilde{R}$ to $\tilde{U}$.
So far we have not done much: 
the pair $\tilde{R}, \tilde{U}$ is an alternate atlas and relations for the {\em same} moduli space ${\cal M}$:
\begin{equation}
\tilde{U} / \tilde{R} = U/R = {\cal M}.
\end{equation}

Now, in order to get ${\M}$, we replace $\tilde{R}$ by a ${\mathbb Z}_2$ torsor $\overline{R}$ on it, 
defined by the square root of the
ratio of the two holomorphic top-forms
that distinguish $\tilde{R}$ from $R$. In other words,
\begin{equation}
\overline{R} := \left\{h,r,f,g \, | \, (r,f,g) \in \tilde{R}, h^2 =f/g\right\}.
\end{equation}
The ratio $f/g$  itself is a pure number,
and the two values of the square root determiine a ${\mathbb Z}_2$ torsor.  The quotient stack
\begin{equation}
\M := \tilde{U} / \overline{R}
\end{equation}
is a gerbe over $U/R =  {\cal M}$, on which the pullback of the Hodge line bundle has
a canonical square root. This is our second definition of the Bagger-Witten line bundle.

\noindent {\bf Remark} These constructions are of course very general. 
For any positive integer $n$ and any line bundle ${\cal L}_H$ on any space ${\cal M}$,
there is an obvious construction of 
a ${\mathbb Z}_n$-gerbe  $p: \M \to {\cal M}$ over ${\cal M}$ 
and a natural $n$-th root of the pullback ${{\overline{{\cal L}_H}}} := p^*({\cal L}_H)$ of ${\cal L}_H$ to $\M$.

\subsubsection{A more abstract definition via classifying spaces} \label{classifying}

 It is natural to reformulate the previous definitions in terms of classifying spaces. 
 
 Let $B\mathbb{C}^*$ be the classifying space of $\mathbb{C}^*$. 
 It has a universal line bundle ${\cal L}$ which generates its Picard group: 
 $\rm{Pic}(B\mathbb{C}^*) = \mathbb{Z}$ and ${\cal L}$ corresponds to the positive generator
 $1 \in \rm{Pic}(B\mathbb{C}^*) = \mathbb{Z}$. Let $f: {\cal M} \rightarrow B\mathbb{C}^*$ denote the classifying map of ${\cal L}_H$ (any line bundle over any stack ${\cal X}$ determines a morphism $g: {\cal M} \rightarrow B\mathbb{C}^*$, called its classifying map, such that $g^*{\cal L} \cong \gamma$).

\noindent The central extension 
\begin{equation}
\begin{tikzcd}
1 \ar[r] & \mathbb{Z}_2 \ar[r] & \mathbb{C}^* \ar[r, "(-)^2"] & \mathbb{C}^* \ar[r] & 1
\end{tikzcd}
\end{equation}
defines a $\mathbb{Z}_2$ gerbe $p: B\mathbb{C}^* \rightarrow B\mathbb{C}^*$. Applying $\rm{Hom}(-, {\mathbb{C}}^*)$ to this sequence, one sees that the induced map $\rm{Pic}(B\mathbb{C}^*) \rightarrow \rm{Pic}(B\mathbb{C}^*)$ is multiplication by 2, so that $p^*{\cal L} \cong {\cal L}^{\otimes 2}$. Furthermore, an arbitrary line bundle $\gamma$ over a stack ${\cal X}$ admits a square root over ${\cal X}$ if and only if the classifying map $g: {\cal X} \rightarrow B\mathbb{C}^*$ factors through the gerbe $p$. We can now define ${\M}$ to be the fiber product: 
\begin{equation}
\M:=  {\cal M} \times_{B\mathbb{C}^*,p} B\mathbb{C}^*.
\end{equation}

\noindent The canonical map $p_{{\cal M}}: \M \rightarrow {\cal M}$ is a $\mathbb{Z}_2$-gerbe, since it is the base change of a $\mathbb{Z}_2$ gerbe. The pullback of the universal bundle ${\cal L}$ along the other projection $\M \rightarrow B\mathbb{C}^*$ is a square root of $\pi^* {\cal L}_H$; this is our third definition of the Bagger-Witten line bundle ${\cal L}_{\rm{BW}}$. 

\noindent {\bf Remark} 
It follows from the previously mentioned properties of $p$ that an arbitrary morphism $g: {\cal X} \rightarrow {\cal M}$ factors through the gerbe $\pi$ if and only if $g^*\lambda$ admits a square root over ${\cal X}$. In particular, the gerbe $\M$ is trivial over ${\cal M}$ if and only if ${\cal L}_H$ admits a square root (in that case, the previous property yields a trivializing section ${\cal M} \rightarrow \overline{\cal M}$ of the gerbe).

 \subsubsection{Equivalence of the definitions} \label{proof}

We verify that the three descriptions of $\overline{\cal M}$ and the Bagger-Witten bundle over it in subsection~\ref{root} (via roots), in subsection~\ref{atlas} (via atlases and relations) and in subsection~\ref{classifying} (via classifying spaces) agree with each other. For the purposes of this section, we let $\overline{\cal M}$ denote the classifying space definition, and prove that it satisfies the first two descriptions. 

Note that an object of $\overline{\cal M}$ (according to its classifying space definition) over the point $\rm{Spec} \ \mathbb{C}$ is by definition a Calabi-Yau $X$, a complex line $V$ (i.e. a line bundle over $\rm{Spec} \ \mathbb{C}$), and an isomorphism $H^0(X, K_X) \cong V^{\otimes 2}$. The last piece of data is equivalent by adjunction to an isomorphism $K_X \cong {\cal V}^{\otimes 2}$, where ${\cal V}$ is a trivializable line bundle over $X$ (precisely, the pullback of $V$ along $X \rightarrow {\rm Spec} \ \mathbb{C}$). This description evidently agrees with our original definition of $\overline{\cal M}$ in subsection~\ref{root}. Now, the fiber of ${\cal L}_{BW}$ over $(X, V^{\otimes 2} \cong H^0(X,K_X))$ is by definition $V$, which is equal to $H^0(X, \cal{V})$, agreeing with the description of the Bagger-Witten line bundle in subsection~\ref{root}. 

We give a general construction of an atlas for $\overline{\cal M}$ that will specialize to that given in subsection~\ref{atlas}. Let $q: U \rightarrow {\cal M}$ denote an atlas, and $R = U \times_{{\cal M}} U$ the relations. If we make the additional hypothesis that $q^*{\cal L}_H$ admits a square root over $U$, then a choice of such a square root defines a morphism $\bar{q}: U \rightarrow \overline{\cal M}$ such that $p\bar{q} = q$ (see the remark in subsection~\ref{classifying}). Actually, $\bar{q}$ fits into a diagram
\begin{equation}
\begin{tikzcd}
U \times_{{\cal M}} \overline{\cal M} \ar[r] \ar[d] & \overline{\cal M} \ar[d, "\pi"] \\
U \ar[r, "q",swap] \ar[bend left=30, u, "s"] \ar[ru, "\bar{q}"] & {\cal M}.
\end{tikzcd}
\end{equation}
The section $s$ trivializes the gerbe $U \times_{{\cal M}} \overline{\cal M}$, giving an isomorphism $U \times_{{\cal M}} \overline{\cal M} \cong U \times B\mathbb{Z}_2$ over $U$. Since the natural map $U \rightarrow U \times B\mathbb{Z}_2$ is representable, and $U \times_{{\cal M}} \overline{\cal M} \rightarrow \overline{\cal M}$ is representable as the base change of the representable map $q$, we conclude that $\bar{q}$ is representable. Hence $U$ is also an atlas for the stack $\overline{\cal M}$ with relations $U \times_{\overline{\cal M}} U$, which is a $\mathbb{Z}_2$ torsor over $R$. 

The atlas construction above fails if $q^*{\cal L}_H$ does not admit a square root over $U$. However, we can always replace $U$ with $\tilde{U}$, the complement of the zero section in the total space of $q^*{\cal L}_H$. Of course, the pullback of $q^*{\cal L}_H$ to $\tilde{U}$ has a canonical trivialization $\mathcal{O}_{\tilde{U}} \cong q^* \mathcal{L}_H$ (induced by the global section $(u,\omega) \mapsto \omega$ of $q^*{\cal L}_H$), so the construction of the previous paragraph can be applied instead to the new atlas $\tilde{q}: \tilde{U} \rightarrow {\cal M}$ (with relations $\tilde{R} = \tilde{U} \times_{{\cal M}} \tilde{U}$). The isomorphism ${\cal O}_{\tilde{U}}^{\otimes 2} \cong {\cal O}_{\tilde{U}} \cong q^*{\cal L}_H$ defines a square root of $q^*{\cal L}_H$, hence a morphism $\tilde{U} \rightarrow \overline{\cal M}$. Explicitly, $\bar{q}$ maps $(u,\omega) \in \tilde{U}$ to the Calabi-Yau $X_u$, together with the square root $\mathbb{C}^{\otimes 2} \cong (q^*\mathcal{L}_H)_{(u,\omega)} = H^0(X_{u}, K_{X_u})$ (here, $(q^*\mathcal{L}_H)_{(u,\omega)}$ refers to the fiber of this line bundle over $(u,\omega) \in \tilde{U}$). We claim that $\tilde{U} \times_{\overline{\cal M}} \tilde{U}$ is isomorphic to $\bar{R}$ from subsection~\ref{atlas}. Explicitly, a point $\mathfrak{p}$ of $\tilde{R}$ consists of points $u,u' \in U$, forms $\omega \in H^0(X_u, K_{X_u})$, $\omega' \in H^0(X_{u'}, K_{X_{u'}})$, and an isomorphism $f: X_u \rightarrow X_{u'}$ such that $f^*\omega' = \omega$. A point $\mathfrak{q}$ in the fiber of $\bar{R} \rightarrow \tilde{R}$ over $\mathfrak{p}$ consists of the same data, together with an isomorphism $\phi: \mathbb{C} \cong \mathbb{C}$ (between the given square roots of $H^0(X_u, K_{X_u})$ and $H^0(X_{u'}, K_{X_{u'}})$, respectively) such that the diagram
\begin{equation}
\begin{tikzcd}
\mathbb{C}^{\otimes 2} \ar[d, "\phi"] \ar[r] & H^0(X_{u'}, K_{X_{u'}}) \ar[d, "f^*"] \\
\mathbb{C}^{\otimes 2} \ar[r] & H^0(X_{u}, K_{X_{u}})
\end{tikzcd}
\end{equation}
commutes. In other words, we have to give $\lambda \in \mathbb{C}$ such that $\lambda^2 = f^*\omega'/\omega$, agreeing with the first description of $\bar{R}$. In conclusion, we have shown that $\overline{\cal M}$ admits an atlas as described in subsection~\ref{atlas}, and proven the equivalence of all definitions.

\subsubsection{Relation to spinors} \label{spinors}

Everything so far in this section was quite general. In particular, there was no reason to consider only square (as opposed to $n$-th order) roots. On the other hand, only square roots appear in the description in terms of spinors in the previous section~\ref{intuition}. We would like to note here that the spinorial construction is indeed a special case of our current approach. 

We return to the general setup of subsection~\ref{classifying}, letting ${\cal M}$ denote an analytic stack and ${\cal L}$ a holomorphic line bundle over it. However, we will assume that ${\cal L}$ is of the form ${\cal L} = \det{\cal V}$, where ${\cal V}$ is a rank $r$ complex vector bundle over ${\cal M}$. Once again, $\M = {\cal M} \times_{B\mathbb{C}^*} B\mathbb{C}^*$ denotes our $\mathbb{Z}_2$ gerbe over $\cal M$, and $p_{\cal M}: {\M} \rightarrow {\cal M}$ denotes the canonical map. We have seen that $p_{\cal M}^*{\cal L}$ admits a holomorphic square root $\mathcal{K}$ over $\M$. Hence, we have that $c_1(p_{\cal M}^*{\cal L}) = 2 c_1({ \cal K})$, and thus $c_1(p_{\cal M}^*{\cal V}) = c_1(\det(p_{\cal M}^*{\cal V})) = c_1(p_{\cal M}^*{\cal L}) = 2c_1({\cal K})$. Reducing mod 2, we find that $w_2(p_{\cal M}^*{\cal V}) = 0$, and hence the pullback ${\cal W} := p_{\cal M}^*\cal{V}$ admits a spin structure. This spin structure, of course, is not uniquely determined: the set of isomorphism classes of spin structures on ${\cal W}$ is a torsor over $H_1(\M, \mathbb{Z}_2)$, and so if $\M$ is not simply connected we should not expect the spin structure to be unique. 

We can specialize again to $\cal{M}$ being a moduli stack of (polarized) Calabi-Yau manifolds. Applying the discussion above to the relative cotangent bundle $\Omega^1_{{\cal U}/{\cal M}}$ of the universal Calabi-Yau ${\cal U} \rightarrow {\cal M}$ over the moduli space, we conclude that its pullback to $\M$ admits a spin structure (as $\M$ was constructed by applying subsection~\ref{classifying} to the determinant of $\Omega^1_{{\cal U}/{\cal M}}$). Let $P \rightarrow \M$ denote the principal spin bundle constructed from the chosen spin structure on the pullback of $\Omega^1_{{\cal U}/{\cal M}}$, $S$ the spinor representation of ${\rm Spin} (2n)$ (where $n$ is the dimension of a Calabi-Yau appearing in $\cal M$), and ${\cal S}$ the associated complex vector bundle over $\M$. 

For any Calabi-Yau manifold $X$, we have a two dimensional space of covariantly constant spinors. If we assume furthermore that $\rm{dim}(X)$ is odd, then the space of covariantly constant spinors breaks up canonically as the direct sum of two complex lines: the (covariantly constant) positive and negative half-spinors (or: the chiral and anti-chiral spinors). Working over the universal Calabi-Yau $\overline{{\cal U}} \rightarrow \M$, and assuming that the Calabi-Yau manifolds appearing in ${\cal M}$ have odd complex dimension, we then have a line subbundle ${\cal L}_{BW}$ of $\cal S$ with fiber over any given $X$ the complex line of positive parallel half-spinors on $X$. 

We have a $\mathbb{C}$-bilinear pairing 
\begin{equation}
{\cal S} \otimes {\cal S} \: \longrightarrow \: {\rm Cliff}\left( p_{\cal M}^*\Omega^1_{{\cal U}/{\cal M}}\right) \: \cong \: {\Lambda}^{\bullet}\left(p_{\cal M}^*\Omega^1_{{\cal U}/{\cal M}} \right)
\end{equation}
(this is the Clifford pairing, implemented in components by gamma matrices). The line bundle $p_{\cal M}^*{\cal L}_H$ embeds into ${\Lambda}^{\bullet}(p_{\cal M}^*\Omega^1_{{\cal U}/{\cal M}})$, where again ${\cal L}_H$ is the Hodge bundle. The subbundle ${\cal L}_{BW}$ thus pairs with itself non-trivially and maps to $p_{\cal M}^*{\cal L}_H$ (the Clifford pairing is non-degenerate), furnishing a square root of the Hodge bundle.

Hence, we find that our gerbe $\M$ carries a line bundle of covariantly constant spinors squaring to the pullback of ${\cal L}_H$, matching the description in section~\ref{intuition}.

 \subsubsection{Elliptic curves} \label{metaplectic}

We can now explicitly describe $\M$ in the case that ${\cal M} = {\cal M}_{1,1}$ is the moduli stack of elliptic curves. In this case, the (analytic) atlas 
\begin{equation}
U = \mathfrak{h} \: \longrightarrow \: {\cal M}_{1,1} \cong [\mathfrak{h}/SL(2,  \mathbb{Z})]
\end{equation}
(where $\mathfrak{h}$ denotes the upper half plane) will suffice. Consider the metaplectic stack $[\mathfrak{h}/Mp(2,   \mathbb{Z})]$, which is a $\mathbb{Z}_2$ gerbe over ${\cal M}_{1,1}$ via the quotient $Mp(2,   \mathbb{Z}) \rightarrow SL(2,  \mathbb{Z})$. The pullback of ${\cal L}_H$ to the metaplectic stack admits a square root, which we describe now. Firstly, recall the following explicit description of $Mp(2,\mathbb{Z})$:
\begin{equation}
Mp(2,\mathbb{Z}) \: = \: \left\{(g,\eta) \in SL(2,\mathbb{Z}) \times \Gamma\left(
{\mathfrak h}, {\cal O}^*_{\mathfrak{h}}\right)\ | \ \eta(\tau)^2 = c\tau + d \right\},
\end{equation}
where as usual,
$$
g = \left( \begin{array}{cc}
a & b \\ c & d \end{array} \right).
$$
The group law is $(g_1,\eta_1) \cdot (g_2, \eta_2) = (g_1 g_2, \eta)$, where $\eta(\tau) = \eta_1(g_2(\tau))\eta_2(\tau)$ (any $\tau \in \mathfrak{h}$). A line bundle on $[\mathfrak{h}/Mp(2,\mathbb{Z})]$ consists of a line bundle ${\cal S}$ over $\mathfrak{h}$ together with descent data, meaning a lift of the action of $Mp(2,\mathbb{Z})$ on $\mathfrak{h}$ to ${\cal S}$. We take ${\cal S} = {\cal O}_\mathfrak{h} \sqrt{dz}$ and define an action of $Mp(2,\mathbb{Z})$ on this sheaf by $(g,\eta) \cdot \sqrt{dz} = \eta^{-1} \sqrt{dz}$. It is clear that this line bundle, which we call ${\cal L}^{1/2}$, is a square root of the pullback of the Hodge bundle (the latter is defined by the invertible sheaf ${\cal O}_{\mathfrak{h}}dz$ on $\mathfrak{h}$, where $(g, \eta) \cdot dz = (c\tau + d)^{-1}dz$). 

Hence, we obtain (see the remark in subsection~\ref{classifying}) a morphism $F: [\mathfrak{h}/Mp(2,   \mathbb{Z})] \rightarrow \overline{\cal M}$ over ${\cal M}_{1,1}$. Let $R'$ denote the relations of $[\mathfrak{h}/Mp(2,   \mathbb{Z})]$. The map $F$ induces of a morphism $[\mathfrak{h} \rightrightarrows \bar{R}] \rightarrow [\mathfrak{h} \rightrightarrows R']$ of the presenting groupoids over ${\cal M}_{1,1}$. The induced map on the atlas $\mathfrak{h}$ is the identity, so $F$ is determined by the induced morphism $R' \rightarrow \bar{R}$ of $\mathbb{Z}_2$ torsors over $\tilde{R}$. We will show that this morphism induces a bijection on closed points, hence is an isomorphism. 

\noindent We have
\begin{equation}
\bar{R}(\mathbb{C}) \: = \: \left\{(\tau, \tau', g, \lambda) \in \mathfrak{h} \times \mathfrak{h} \times SL(2,\mathbb{Z}) \times \mathbb{C}^* \ | \ g\tau = \tau', \lambda^2 = (c\tau + d) \right\},
\end{equation}
and,
\begin{equation}
R'(\mathbb{C}) \: = \: \left\{(\tau,\tau',g,\eta) \in \mathfrak{h} \times \mathfrak{h} \times SL(2,\mathbb{Z}) \times \Gamma\left( {\mathfrak h},
{\cal O}^*_{\mathfrak{h}}\right) \ | \ g\tau = \tau', \eta(z)^2 = (cz + d)\right\}.
\end{equation}
The induced map $R'(\mathbb{C}) \rightarrow \bar{R}(\mathbb{C})$ evaluates the function $\eta$ at $\tau \in \mathfrak{h}$, and is a bijection (the function $cz + d$ has two square roots over $\mathfrak{h}$, and these are never equal at a point). Therefore $R' \rightarrow \bar{R}$ is an isomorphism, and the same is true for $F$. By construction of $F$, $F^*{\cal L}_{BW}$ is the chosen square root ${\cal L}^{1/2}$ of the Hodge bundle over $[\mathfrak{h}/Mp(2,\mathbb{Z})]$.

\subsection{Resolution of a puzzle}

So far we have described Bagger-Witten line bundles in terms of
covariantly constant spinors.  However, as previously noted,
there are multiple covariantly constant spinors -- two on a generic
Calabi-Yau threefold.  Witten originally defined the Bagger-Witten line
bundle in terms of holomorphic transition functions $\exp(-f/2)$, 
determined by
the K\"ahler potential as
\begin{equation}
K \: \mapsto \: K \: + \: f \: + \: \overline{f}.
\end{equation}
Consistency would appear to require an analogous ambiguity or symmetry in the
definition above.  Based on our statements above, for a generic
Calabi-Yau threefold, we expect that the two Bagger-Witten line bundles
are ${\cal L}_{\rm BW}$ and ${\cal L}_{\rm BW}^{-1}$, suggesting that the
supergravity theory should be invariant under $f \mapsto -f$, which is
not at all clear.

This puzzle is resolved by the work of \cite{Gomis:2015yaa},
who as interpreted in \cite{mebw}, argued that the Hodge and Bagger-Witten
line bundles should be flat\footnote{
At least for moduli spaces of maximal holonomy Calabi-Yau $n$-folds for
$n \neq 2$.
}.  For a flat bundle, we can choose transition
functions $\exp(-f/2)$ to be constant (and in this case, $f$'s pure imaginary).
The four-dimensional supergravity is clearly invariant under $f 
\leftrightarrow \overline{f}$, and under that
action, since for $f$ pure imaginary, $\overline{f} = - f$, 
we see that both ${\cal L}_{\rm BW}$
and ${\cal L}_{\rm BW}^{-1}$ can act as Bagger-Witten line bundles
in the sense above, as expected.

It is worth repeating that this is consistent with known examples.
For moduli spaces of elliptic curves \cite{mebw}, both Hodge and Bagger-Witten
line bundles are nontrivial and flat.
The moduli spaces of toroidal orbifolds discussed in
\cite{ds17} and in section~\ref{sect:toroidalexs} are similar:
the Hodge
line bundles are, again, flat, generating finite subgroups of the Picard
group, hence the Bagger-Witten line bundles are also flat,
since they are square roots of the Hodge line bundle.

Conversely, we know that Calabi-Yau manifolds always have (at least) two
covariantly constant spinors, and if the symmetry exchanging them in the
supergravity theory is $f \mapsto \overline{f}$, then this would appear
to require that $\overline{f}$ be pure imaginary, suggesting that
Bagger-Witten line bundles must always be flat, an alternative to
the argument in \cite{Gomis:2015yaa}.
Since Bagger-Witten line bundles are square roots of Hodge line bundles,
this would imply that Hodge line bundles must also always be flat.
It would be interesting to understand the Bagger-Witten and Hodge line
bundles in a greater variety of examples, to better appreciate
subtleties.  (See also \cite{Donagi:2017vwh} for some related
observations.)

\section{Some warm-up models}
\label{sect:warmupmodels}

Before describing the Bagger-Witten line bundles over the
moduli spaces of toroidal orbifolds constructed in
\cite{ds17}, first we will warm-up with some models that are
easier to work with explicitly. 
Specifically, we will construct Bagger-Witten line bundles on
(${\mathbb Z}_2$ gerbes on) moduli spaces that are products
of the spaces 
\begin{equation}
{\cal M}(2) \: \equiv \: [ {\mathfrak h} / \Gamma(2) ]
\end{equation}
and 
\begin{equation}
{\cal M}_1(2) \: \equiv \: [ {\mathfrak h} / \Gamma_1(2) ], 
\end{equation}
where ${\mathfrak h}$ is the upper half plane, and
$\Gamma(2)$ and $\Gamma_1(2)$ are (congruence) subgroups of
$SL(2,{\mathbb Z})$.  These are
moduli spaces of
elliptic curves with level structures, and we have outlined their properties
in
appendix~\ref{app:picard}.  Some such moduli spaces appeared in the work
\cite{Hebecker:2017lxm}, as moduli spaces of toroidal orientifolds
with flux.  
Although we do not know of Calabi-Yau manifolds whose moduli spaces
correspond to every possible product of ${\cal M}(2)$, 
${\cal M}_1(2)$, their relative ease-of-use and explicit description
nevertheless make them handy for technical illustrations of the construction of
Bagger-Witten line bundles, which is what we shall do in this section.

Before going further, let us also perform a simple consistency check
of these models.
In \cite{Ooguri:2006in}, 
it was proposed that a necessary condition for existence of UV
completion in a gravity theory is that
the closure of the moduli space should
be simply connected.
(See also {\it e.g.}
\cite{Brennan:2017rbf}[section 2.7], \cite{ArkaniHamed:2006dz}.)
We are starting with stacky moduli `spaces'
 ${\cal M}(2)$ and ${\cal M}_1(2)$,
keeping track of $SL(2,{\mathbb Z})$
actions (and actions of certain subgroups),
and also remembering trivially-acting group actions.  
These stacky moduli spaces are not simply-connected, but the
varieties underlying ${\cal M}(2)$ and ${\cal M}_1(2)$ both compactify
to\footnote{
In more detail, the compactification of the variety underlying
${\cal M}(2)$, sometimes denoted $X(2)$,
is a ${\mathbb P}^1$ with coordinate $\lambda$ and with a universal
curve
\begin{equation}
y^2 \: = \: x(x-1) (x - \lambda).
\end{equation}
The symmetric group $S_3$ acts on $X(2)$ sending $\lambda$ to the six
cross-ratios of the four points $0$, $1$, $\lambda$, $\infty$,
and the quotient $X(2)/S_3$ is the coarse moduli space of elliptic
curves.  The quotient $X(2)/{\mathbb Z}_2$, for ${\mathbb Z}_2 \subset
S_3$, is $X_1(2)$, the coarse moduli space underlying ${\cal M}_1(2)$.
See also e.g. \cite{ogg}.
} genus zero curves.  Therefore, in terms of the underlying varieties,
omitting the stack structure,
the closure of the moduli spaces in these examples is simply-connected,
consistent with the proposal \cite{Ooguri:2006in} that closures
of moduli spaces be simply-connected.

Now, let us turn to the construction of Bagger-Witten line bundles.
As explained in appendix~\ref{app:picard}, much as in the case of
ordinary elliptic curves, the Hodge line bundle is well-defined
over both ${\cal M}(2)$ and ${\cal M}_1(2)$, but it does not have
a square root there.  In both cases, as well as $[ {\mathfrak h} /
SL(2,{\mathbb Z}) ]$, the Hodge line bundle is nontrivial and generates
the Picard group.  If the entire moduli space is a single
${\cal M}(2)$ or ${\cal M}_1(2)$, then we need to construct take a local
trivially-acting group quotient (technically, aa
${\mathbb Z}_2$ gerbe), over which the Hodge line bundle does admit
a square root.

Some natural choices of such ${\mathbb Z}_2$ gerbes are outlined
in appendix~\ref{app:picard}.  Specifically, following the same model
as for ordinary elliptic curves, we can pull back the congruence
subgroups $\Gamma(2), \Gamma_1(2) \subset SL(2,{\mathbb Z})$ to
$Mp(2,{\mathbb Z})$.  If we let $p: Mp(2,{\mathbb Z}) \rightarrow
SL(2,{\mathbb Z})$ denote the projection, then instead of
${\cal M}(2)$ or ${\cal M}_1(2)$, we could consider instead
\begin{equation}
[ {\mathfrak h} / p^* \Gamma(2) ], 
\: \: \:
[ {\mathfrak h} / p^* \Gamma_1(2) ],
\end{equation}
respectively.  As discussed in appendix~\ref{app:picard}, for these
stacks (which are ${\mathbb Z}_2$ gerbes over ${\cal M}(2)$,
${\cal M}_1(2)$), the Hodge line bundle is the square of a generator
of the Picard group.  Hence, we can identify each Bagger-Witten line bundle
with a generator.

If the moduli space is a product of several factors of the form above,
then the construction of the ${\mathbb Z}_2$ gerbe is more subtle.
Naively, one might imagine taking a product of factors in which each
factor has a ${\mathbb Z}_2$ gerbe -- but that goes too far, providing
more roots than just the single square root actually required.
The construction in this case is more subtle, as we describe next.

Consider for example a product of three factors of the form
\begin{equation}
[ {\mathfrak h}/\Gamma_A] \times
[ {\mathfrak h}/\Gamma_B] \times
[ {\mathfrak h}/\Gamma_C],
\end{equation}
where ${\mathfrak h}$ is the upper half plane and
$\Gamma_{A, B, C}$ are each equal to either $\Gamma(2)$ or
$\Gamma_1(2)$.  
In each case, the Hodge line bundle corresponds to an element of the
Picard group of the product of the form $(g_A, g_B, g_C)$,
where each $g$ generates the Picard group of the factor.

Now, the moduli space for which the Bagger-Witten line bundle is
defined
is a ${\mathbb Z}_2$ gerbe over the moduli spaces above, as it
is a square root of the Hodge line bundle.  (In particular, we go to
a ${\mathbb Z}_2$ gerbe over the original moduli spaces, because the
Hodge line bundle admits no square roots over the original moduli spaces
but may over a suitable ${\mathbb Z}_2$ gerbe.)
We construct it explicitly as follows.

First, for the $({\mathbb Z}_2)^3$ appearing as the kernel
\begin{equation}
1 \: \longrightarrow \: ({\mathbb Z}_2)^3 \: \longrightarrow \:
Mp(2,{\mathbb Z})^3 \: \longrightarrow \: SL(2,{\mathbb Z})^3
\: \longrightarrow \: 1,
\end{equation}
define $f: ({\mathbb Z}_2)^3 \rightarrow {\mathbb Z}_2$ to be the
product of the elements of $({\mathbb Z}_2)^3$,
and define $K$ to be the kernel of $f$.

Then, define
\begin{equation}
A_1 \: \equiv \: Mp(2,{\mathbb Z})^3 / K.
\end{equation}
The map $Mp(2,{\mathbb Z})^3 \rightarrow SL(2,{\mathbb Z})^3$ descends
to a map $A_1 \rightarrow SL(2,{\mathbb Z})^3$, with kernel
${\mathbb Z}_2$:
\begin{equation}
1 \: \longrightarrow \: {\mathbb Z}_2 \: \longrightarrow \:
A_1 \: \longrightarrow \: SL(2,{\mathbb Z})^3 \: \longrightarrow \: 1.
\end{equation}

Finally, define
\begin{equation}
A_2 \: \equiv \: A_1 \times_{ SL(2,{\mathbb Z})^3 }
(\Gamma_A \times \Gamma_B \times \Gamma_C).
\end{equation}
By construction of the fiber product,
there are natural maps from $A_2$ to both $A_1$ and
$\Gamma_A \times \Gamma_B \times \Gamma_C$.
The map
\begin{equation}
A_2 \: \longrightarrow \:
\Gamma_A \times \Gamma_B \times \Gamma_C
\end{equation}
is surjective with kernel ${\mathbb Z}_2$:
\begin{equation}
1 \: \longrightarrow \: {\mathbb Z}_2 \: \longrightarrow \: A_2 \:
\longrightarrow \:
\Gamma_A \times \Gamma_B \times \Gamma_C \: \longrightarrow \: 1.
\end{equation}

Finally, the ${\mathbb Z}_2$ gerbe over the moduli space, over which
square roots of the Hodge line bundle exist, is given by
\begin{equation}
[ {\mathfrak h}^3 / A_2 ],
\end{equation}
which is a ${\mathbb Z}_2$ gerbe over
\begin{equation}
[ {\mathfrak h}/\Gamma_A ] \times
[ {\mathfrak h}/\Gamma_B ] \times
[ {\mathfrak h}/\Gamma_C ].
\end{equation}

In particular, if we had needed to define the Bagger-Witten
line bundle for each factor separately,
then we would need a moduli space of the
form
\begin{equation}
[ {\mathfrak h} / p^* \Gamma_A ] \times
[ {\mathfrak h} / p^* \Gamma_B ] \times
[ {\mathfrak h} / p^* \Gamma_C ],
\end{equation}
where $p: Mp(2,{\mathbb Z}) \rightarrow SL(2,{\mathbb Z})$.
However, because the three Fock vacua of the individual tori
are always multiplied together, in the form
\begin{equation}
| \pm \rangle_1 \otimes | \pm \rangle_2 \otimes | \pm \rangle_3,
\end{equation}
we do not need to define them
separately to build a moduli space of SCFTs -- we only need for the
product of the three vacua to be well-defined across the moduli space.
This is why we take
an extension by a single ${\mathbb Z}_2$, defined by the
product map $f$ above, rather than a metaplectic extension of
each individual factor.

Before moving on, we should address the physical meaning of these
extra ${\mathbb Z}_2$ gerbe structures which we are forced to introduce in
order to define the Bagger-Witten line bundle.  In the case of
moduli spaces of elliptic curves, we argued in
\cite{meta} that this structure was physically meaningful,
encoding an extra ${\mathbb Z}_2$ T-duality, and also arose in
related string dualities.  In principle, the same statement is also true
in more general cases:  an extra ${\mathbb Z}_2$ involution in the
definition of the moduli space of SCFTs implicitly implies an extra
${\mathbb Z}_2$ self-duality in the target space low-energy theory, 
and for the ${\mathbb Z}_2$
arising in defining the Bagger-Witten line bundle, given its role in
four-dimensional supergravities, the implication is that the
extra ${\mathbb Z}_2$ self-duality acts on fermions.

\section{Three-dimensional toroidal orbifolds}
\label{sect:toroidalexs}

Next, we will apply our construction to the moduli spaces of toroidal
orbifolds with $h^{2,1}=3$ obtained in \cite{ds17}. We briefly recall the setup and results of that paper here. Let $E_i$ ($i=1,2,3$) denote a triple of elliptic curves, $Y = E_1 \times E_2 \times E_3$. We consider a finite group $G = (\mathbb{Z}_2)^{r+2}$ ($0 \leq r \leq 6$) acting on $Y$ by a combination of elliptic involutions (i.e. $z_i \mapsto -z_i$) and translations by points of order 2: the precise group actions considered are described in the appendix of \cite{ds17}, as well as Table 1 of \cite{dw}. The basic example the reader should keep in mind is $G = (\mathbb{Z}_2)^2$, acting on $Y$ by an even number of elliptic inversions (e.g., $(z_1, z_2, z_3) \mapsto (-z_1, -z_2, z_3)$). The quotient $X = Y/G$ is in all cases a possibly singular complex orbifold, but admits a crepant resolution $\overline{X} \rightarrow X$ (described in section 5 of \cite{ds17}). $\overline{X}$ is a Calabi-Yau manifold of dimension 3, and its complex structure moduli space is known explicitly for 10 of the listed $G$ actions.

Firstly, note that $H_{\rm max} := (\mathbb{Z}_4)^6 \rtimes \Gamma^3 \rtimes S_3$ (where as usual $\Gamma := SL(2,\mathbb{Z})$ and $S_3$ is the symmetric group on three letters) acts on $\mathfrak{h}^3$ ($(\mathbb{Z}_4)^6$ acts trivially, $\Gamma^3$ acts componentwise by fractional linear transformations, and $S_3$ acts to permute the three factors of $\mathfrak{h}$). We have an embedding of $G$ into $H_{\rm max}$. Its normalizer in $H_{\rm max}$ is denoted $H$, and the quotient is $H' := H/G$. Over $\mathfrak{h}^3$, there is a family ${\cal Y} \rightarrow \mathfrak{h}^3$ of abelian varieties, and $G$ acts on this family to yield a deformation ${\cal Y}/G \rightarrow \mathfrak{h}^3$ of $X$. The classifying map of this family, $\mathfrak{h}^3 \rightarrow {\cal M}_X$ (to the moduli space of $X$), then induces an isomorphism ${\cal M}_X \cong [\mathfrak{h}^3/H']$ in the relevant cases (when $h^{2,1}(X) = 3$). Lastly, as the particular resolution $\overline{X} \rightarrow X$ can be performed in families, we identify ${\cal M}_{\overline{X}}$ with ${\cal M}_X$.

We want to construct our gerbe $\M$ over ${\cal M } := {\cal M}_{\overline{X}}$ and identify the Bagger-Witten bundle on it. We will use the atlas and relations description provided in subsection~\ref{atlas}. Our atlas is $U = \mathfrak{h}^3$. The Hodge bundle on ${\cal M}$ has a simple description: we begin with the invertible sheaf ${\cal O}_{\mathfrak{h}^3}[dz_1 \wedge dz_2 \wedge dz_3]$, and define an action of $(a,g,\sigma)$ (here, $a \in (\mathbb{Z}_2)^6$, $g = (g_1, g_2, g_3) \in \Gamma^3$, and $\sigma \in S_3$) on this sheaf by
$$
(a,g,\sigma) \cdot dz_1 \wedge dz_2 \wedge dz_3 \: = \: (-1)^{|\sigma|}(c_1\tau_1 + d_1)^{-1}(c_2\tau_2 + d_2)^{-1}(c_3\tau_3 + d_3)^{-1} dz_{1} \wedge dz_{2} \wedge dz_{3}
$$
where
$$
g_i = \left( \begin{array}{cc}
a_i & b_i \\ c_i & d_i \end{array} \right),
$$
and $|\sigma| \in \mathbb{Z}_2$ is the sign of $\sigma \in S_3$. We start with some group theoretic setup: define a group $\tilde{H}_{{\rm max}}$ by
$$
\tilde{H}_{{\rm max}} \: = \: \left\{(h, \eta) \in H_{ \rm max} \times {\cal O}_{\mathfrak{h}^3}^* (\mathfrak{h}^3) \ | \ \eta(\tau_1, \tau_2, \tau_3)^2 = (-1)^{|\sigma|}(c_1\tau_1 + d_1)(c_2\tau_2 + d_2)(c_3\tau_3 + d_3) \right\},
$$
with multiplication $(h,\eta)(h', \eta') = (hh', \eta)$, where $\eta(\tau) = \eta_1(h(\tau))\eta_2(\tau)$ (here we are viewing $h$ as an automorphism of $\mathfrak{h}^3$). The natural projection $\tilde{H}_{\rm max} \rightarrow H_{\rm max}$ has kernel $\mathbb{Z}_2$. We seek a more explicit description of $\tilde{H}_{\rm max}$. Let
\begin{equation}
\tilde{S}_3 \: = \: \left\{(\sigma, \epsilon) \in S_3 \times \mathbb{Z}_4 \ | \ \epsilon^2 = (-1)^{|\sigma|}\right\}.
\end{equation}
An element of $\tilde{S}_3$ is thus an element of $S_3$ together with a square root of its sign. There is a natural map $\tilde{S}_3 \rightarrow S_3$, with kernel $\mathbb{Z}_2$. We can then construct an extension of $H_{\rm max}$ of the form $(\mathbb{Z}_4)^6 \rtimes Mp(2,\mathbb{Z})^3 \rtimes \tilde{S}_3$, where $\tilde{S}_3$ acts on $(\mathbb{Z}_4)^6 \rtimes Mp(2, \mathbb{Z})^3$ via the quotient $\tilde{S}_3 \rightarrow S_3$. There is a map
\begin{equation}
\rho: (\mathbb{Z}_4)^6 \rtimes Mp(2,\mathbb{Z})^3 \rtimes \tilde{S}_3 \:
\longrightarrow \: \tilde{H}_{\rm max}
\end{equation}
defined by 
\begin{equation}
(a, (g_i, \eta_i)_{i = 1,2,3}, \sigma, \epsilon) \: \mapsto \: (h, \epsilon\eta_1\eta_2\eta_3),
\end{equation}
where $h = (a, (g_1,g_2,g_3),\sigma) \in H_{\rm max}$, with $a \in (\mathbb{Z}_4)^6$, $g_i \in SL(2,\mathbb{Z})$, and $\sigma \in S_3$. The map is well-defined since 
\begin{equation}
(\epsilon \eta_1 \eta_2 \eta_3)^2 \: = \: (-1)^{|\sigma|} (c_1\tau_1 + d_1)(c_2\tau_2 + d_2)(c_3 \tau_3 + d_3).
\end{equation}
The kernel $\ker{\rho}$ consists of elements $(0, (1_{SL(2,\mathbb{Z})},\eta_i)_{i=1,2,3}, 1_{S_3}, \epsilon)$ satisfying $\eta_i^2 = \epsilon^2 = 1$ for all $i$, and $\epsilon \eta_1 \eta_2 \eta_3 = 1$ (we let the identity of a group $\Omega$ be denoted $1_{\Omega}$ when there is potential confusion). The first condition defines a subgroup isomorphic to $(\mathbb{Z}_4)^4$ (abstractly: this subgroup should not be confused with the $(\mathbb{Z}_4)^6$ translation subgroup; it intersects it trivially), and the second condition yields a subsequent $(\mathbb{Z}_4)^3$ subgroup. We conclude that 
\begin{equation}  \label{eq:6}
\tilde{H}_{\rm max} \: \cong \: \left((\mathbb{Z}_4)^6 \rtimes Mp(2,\mathbb{Z})^3 \rtimes \tilde{S}_3 \right)/(\mathbb{Z}_4)^3.
\end{equation}
From now on, let $\pi: \tilde{H}_{\rm max} \rightarrow H_{max}$ denote the projection. Let $\tilde{H} = \pi^{-1}(H)$, $\tilde{G} = \pi^{-1}(G)$. There is an isomorphism $\tilde{H}/\tilde{G} \cong H/G = H'$. $G$ itself embeds into $\tilde{G} \leq \tilde{H}$ as a normal subgroup, and we can form the quotient $\tilde{H}' := \tilde{H}/G$.

We return to the construction of the moduli space $\M$, following subsection~\ref{atlas}. Firstly we replace $U = \mathfrak{h}^3$ with $\tilde{U}$, the total space of the punctured Hodge bundle. A point of $\tilde{U}$ is a pair $(\tau, \omega)$, where $\tau = (\tau_1, \tau_2, \tau_3)$ and $\omega$ is a holomorphic top form on $\prod_{i=1}^{3} \mathbb{C}/\langle 1, \tau_i \rangle$. The new space of relations $\tilde{R} = \tilde{U} \times_{\cal M} \tilde{U}$ has points of the form $(\tau, \omega, \tau', \omega', \bar{h})$, where $h \in H$ satisfies $h \cdot \tau = \tau'$, and $\bar{h}$ is the image of $h$ under $H \rightarrow H'$. Next, we construct a $\mathbb{Z}_2$ torsor $\overline{R} \rightarrow \tilde{R}$, with points $(\tau, \omega, \tau', \omega', \bar{h}, \lambda)$, where now $\lambda \in \mathbb{C}^*$ satisfies 
\begin{equation}
\lambda^2 \: = \: h^*\omega'/\omega \: = \: (-1)^{|\sigma|} \prod_{i=1}^{3}(c\tau_i + d_i).
\end{equation}
Then $\M$ is the quotient groupoid $[\tilde{U}/\overline{R}]$.

Note now that $\tilde{H}'$ acts on $\tilde{U}$ by 
\begin{equation}
(\bar{h}, \eta) \cdot (\tau, \omega) \: = \: (h \cdot \tau, \eta \omega)
\end{equation}
and the corresponding space of relations $R'$ has points $(\tau, \omega, \tau', \omega', \bar{h}, \eta)$, with $h \cdot \tau = \tau'$ and 
\begin{equation}
\eta(w_1,w_2,w_3)^2 \: = \: (-1)^{|\sigma|} \prod_{i=1}^{3}(c_iw_i + d_i)
\end{equation}
where we have used $w_i$ to denote coordinates on $\mathfrak{h}^3$ to avoid confusion with the fixed $\tau_i$'s. We have a natural map $R' \rightarrow \bar{R}$ evaluating $\eta$ at $(\tau_1, \tau_2, \tau_3)$ to yield a complex number $\lambda = \eta(\tau)$ satisfying 
\begin{equation}
\lambda^2 \: = \: (-1)^{|\sigma|} \prod_{i=1}^{3}(c\tau_i + d_i).
\end{equation} 
The map is clearly a biholomorphism (if $\eta^2 = \eta'^2$ for nonvanishing $\eta$ and $\eta'$ on $\mathfrak{h}^3$, then $\eta = \pm \eta'$, and thus having $\eta(\tau) = \eta'(\tau)$ for a single $\tau$ would imply $\eta = \eta'$), so we conclude that
\begin{equation}
\M \: = \: [\mathfrak{h}^3/\tilde{H}'].
\end{equation}
The Bagger-Witten bundle then has the following explicit description. Begin with the invertible sheaf ${\cal O}_{\mathfrak{h}^3}\sqrt{dz_1 \wedge dz_2 \wedge dz_3}$, and define an action of $(\bar{h}, \eta) \in \tilde{H}'$ on the generator $\sqrt{dz_1 \wedge dz_2 \wedge dz_3}$ by
\begin{equation}
(\bar{h}, \eta) \cdot \sqrt{dz_1 \wedge dz_2 \wedge dz_3} \: = \: \eta \sqrt{dz_1 \wedge dz_2 \wedge dz_3}.
\end{equation}
This action defines a line bundle ${\cal L}_{BW}$ over $\M$. Since $\eta^2 = (-1)^{|\sigma|}\prod_{i=1}^{3} (c_i\tau_i + d_i)$, we conclude that ${\cal L}_{BW}^{\otimes 2}$ is isomorphic to the pullback of the Hodge bundle, as usual.

Note that in the simple case in which $G = (\mathbb{Z}_2)^2$, and $\overline{X}$ is (a resolution of) the Vafa-Witten orbifold, we have $H = H_{\rm max}$ and thus equation~(\ref{eq:6}) together with $\tilde{H}' = \tilde{H}/G$ gives a quite explicit description of the moduli space.

\section{Criteria for existence of UV completions}
\label{sect:weakgrav}

Over the last few years, there have been several proposals for
criteria to check existence of UV completions of four-dimensional
supergravity theories, such as the weak gravity conjecture 
(see {\it e.g.} \cite{Ooguri:2006in,Brennan:2017rbf,ArkaniHamed:2006dz}).  In this section,
we will propose another, based on properties of the Bagger-Witten line bundle.

First, it has been argued in physics in e.g.
\cite{Gomis:2015yaa} that, over\footnote{
We exclude cases with larger worldsheet symmetries, corresponding for example
to K3 surfaces and to Calabi-Yau $n$-folds whose holonomy is a proper
subgroup of $SU(n)$.
} moduli spaces of maximal holonomy Calabi-Yau
$n$-folds for $n \neq 2$, the Bagger-Witten line bundle is always flat.
(For a more detailed analysis of this statement, and of earlier
work, see \cite{mebw}.)  There are also analogous results
in four-dimensional $N=2$ theories \cite{Niarchos:2018mvl}, namely
that bundles of superconformal primaries over Higgs branches of such
theories admit flat connections. 
 
In fact, in all known examples \cite{mebw,ds17} the
Hodge and Bagger-Witten line bundles are not only flat, but also holomorphically
nontrivial.  Strikingly, from a physics perspective, this includes
the free-field case of elliptic curves -- even there, on a moduli space
of physical theories defined by free fields, the Hodge and Bagger-Witten
line bundles are nontrivial.  Given that they are nontrivial in all known
examples, even for moduli spaces of free field theories, 
we propose the following\footnote{
One of the authors has also previously mentioned this in
\cite{Sharpe:2019yag}.
}
\begin{quotation}
{\it Conjecture:  in any four-dimensional $N=1$ supergravity with a UV
completion, the Bagger-Witten line bundle is holomorphically nontrivial
but with associated flat connection.}
\end{quotation}

This is intended to be a proposal for a 
criterion for UV completions of four-dimensional $N=1$ supergravity theories
specifically.  Certainly for non-supergravity theories, we cannot even
formulate the conjecture, but also for supergravity theories in other
dimensions, there may be other subtleties.
For example, in a compactification to six dimensions,
the $U(1)_R$ worldsheet symmetry
is enhanced (to $SU(2)_R$), which would modify the analysis. 
The conjecture is motivated by perturbative Calabi-Yau compactifications,
but is intended to apply to more general four-dimensional supergravity
theories as well.

In passing, let us also briefly comment on the
simple-connectedness criterion for UV completions.  It was argued in
e.g. \cite{Ooguri:2006in,Brennan:2017rbf} that after projecting
out gerbe structures, compactifications of the moduli spaces
arising in string theory should be simply-connected.
In addition to observing that this worked in examples, 
they also argued that, for example, if one compactifies to $0+1$
dimensions on a spatial circle (amongst other things), then following a
noncontractible loop on the moduli space along that spatial circle
would give a nonzero global charge, contradicting various folk theorems
about existence of UV completions of gravity theories.  We emphasize
that this argument only sees the `untwisted-sector' portion of the
fundamental group of a moduli stack.  For example, a moduli space of
elliptic curves $[ {\mathfrak h} / SL(2,{\mathbb Z} ]$ has 
fundamental group $SL(2,{\mathbb Z})$, including a ${\mathbb Z}_2$ center
that is due to a ${\mathbb Z}_2$ gerbe structure,
but after projecting out gerbe
structures, it compactifies to a genus zero curve.

Based on the statements in the previous section and elsewhere,
we can make a few mathematics conjectures regarding properties of the
Hodge line bundle of holomorphic top forms over a moduli space of
Calabi-Yau manifolds:
\begin{enumerate}
\item We conjecture that
the dual of the Hodge line bundle is ample.
Physically, this follows from an observation in \cite{bw} that
ties properties of the Hodge line bundle to positivity of kinetic
energies.  That said, over noncompact moduli spaces, this is a rather
weak statement, as the notion of 'ample' is comparatively weak,
and over (partial) compactifications, this effectively
acts as a constraint on any extension of the Hodge line bundle over
the compactification divisor, at least for compactifications intended
for physical applications.  
In any event, this conjecture has appeared elsewhere, and we repeat it
here for completeness.
\item We conjecture that
over an uncompactified moduli space of maximal-holonomy Calabi-Yau $n$-folds,
for $n \neq 2$, the Hodge
line bundle is nontrivial but is associated with a flat connection.
This is simply an analogue of the physics conjecture above.
\end{enumerate}

\section{Spacetime superpotential terms}
\label{sect:superpotential:27:3}

As originally noted in \cite{bw}, the four-dimensional $N=1$ supersymmetric
spacetime superpotential is a meromorphic section of 
${\cal L}_{\rm BW}^{\otimes 2} \cong {\cal L}_H$.  It is important to note,
however, that the space over which it lives may be slightly different than
the moduli spaces discussed so far, which resolves a small puzzle as
we will summarize in this section.

So far we have discussed moduli spaces of target-space gauge singlets,
in effect.  In a perturbative heterotic compactification on one of the
Calabi-Yau threefolds discussed in this paper, the `moduli space' over which
the spacetime superpotential is defined would be more than just the
space of singlets -- it would also include vevs of charged matter fields.
For example, in a perturbative heterotic compactification on a Calabi-Yau
threefold on the (2,2)
locus, the target-space theory would contain matter in the
${\bf 27}$ and ${\bf \overline{27}}$ of $E_6$, and the spacetime superpotential
is defined over a space of not only gauge singlets but also
${\bf 27}$ and ${\bf \overline{27}}$ vevs.

We can locally model the structure of that moduli space as follows.
As discussed in \cite{Bershadsky:1993cx}[section 2.3],
for (2,2) supersymmetric worldsheet theories,
the ${\bf 27}$ and ${\bf \overline{27}}$ couple to $T {\cal M} 
\otimes {\cal L}_{\rm BW}$, the tangent bundle of the moduli space twisted
by the Bagger-Witten line bundle.  Therefore, we can locally model
the moduli space over
which the spacetime superpotential is defined to be the total space of
$T {\cal M} \otimes {\cal L}_{\rm BW}$, where the moduli space ${\cal M}$
is the moduli space of gauge singlets discussed elsewhere in this paper.

This resolves a minor puzzle concerning the spacetime superpotential.
For all of the moduli spaces discussed in this paper, both the Bagger-Witten
line bundle and the Hodge line bundle are examples of 'fractional' line bundles on gerbes,
which are not pullbacks of line bundles on underlying spaces.  Such line bundles
on gerbes can have odd properties, such as fractional Chern classes, but more
to the point for our discussion, they do not admit any sections\footnote{
For the moduli spaces appearing in section~\ref{sect:warmupmodels},
we can see this explicitly as follows.
Those moduli spaces 
are products of quotients $[{\mathfrak h}/G]$ where
$G$ is some congruence subgroup of $SL(2,{\mathbb Z})$ or a related group.
A section of the Hodge line bundle would be a modular form for that congruence
subgroup, of degree one.  Such modular forms are discussed in
\cite{diamonds}[section 1.2].  Briefly, a function $f$ is modular for a
congruence subgroup if for any $\gamma \in G$,
\begin{equation}
f(\gamma \tau) \: = \: (c \tau + d)^k f(\tau)
\end{equation}
and $f$ is holomorhpic on ${\mathfrak h}$ and at the cusps (meaning its
transforms are holomorphic at infinity).  In any event, much as for ordinary
modular forms, if $k$ is odd, there are no modular forms of weight $k$
with respect to any congruence subgroup containing $-I$, such as $\Gamma(2)$
and $\Gamma_1(2)$ \cite{diamonds}[section 1.2].
},
holomorphic or meromorphic or smooth.  Therefore, if the spacetime
superpotential were defined over ${\cal M}$, and not the total space
above, then since ${\cal L}_H$ has no sections, the spacetime superpotential
must vanish, contradicting the fact that Yukawa couplings such as
${\bf 27}^3$ and ${\bf \overline{27}}^3$, corresponding to derivatives of
the spacetime superpotential, are nonzero.

The resolution of this puzzle, at least in this model of the (2,2) locus,
lies in the fact that the spacetime superpotential
is defined not over ${\cal M}$, but rather the total space of another
such fractional bundle, namely $T {\cal M} \otimes {\cal L}_{\rm BW}$.
Such a total space is not a gerbe, and so this obstruction to the existence
of the superpotential is removed.

To clarify, let us consider a simple example, the line bundle
sometimes denoted ${\cal O}(-1/2)$ over a ${\mathbb Z}_2$ gerbe on
${\mathbb P}^1$.  We can think of this as the line bundle ${\cal O}(-1)$ on
the weighted projective stack ${\mathbb P}^1_{[2,2]}$, which is realized
in a two-dimensional gauged linear sigma model with gauge group $U(1)$
and fields $x_{0,1}$ with charges
\begin{center}
\begin{tabular}{c|ccc}
& $x_0$ & $x_1$ & $z$ \\ \hline
$U(1)$ & $2$ & $2$ & $-1$
\end{tabular}
\end{center}
We intepret $x_{0,1}$ as homogeneous coordinates on the
${\mathbb P}^1$, and $z$ as a coordinate along the fiber of the line
bundle.  This is a line bundle on the gerbe, but over ${\mathbb P}^1$,
it is merely a fiber bundle, with fibers $[ {\mathbb C} / {\mathbb Z}_2 ]$.
Here we see explicitly in this example that the total space of
a fractional line bundle (a line bundle on a gerbe which is not a pullback
of a line bundle on the base) is not itself a gerbe.

In fact, more can be said.  The zero section of a fractional bundle gives a copy
of the underlying gerbe.  As a result, over the total space of
$T {\cal M} \otimes {\cal L}_{\rm BW}$, the locus where all charged
matter vevs vanish naturally has a gerbe structure -- perfectly reflecting
the physical expectation that that classical enhanced symmetry locus should
be singular.

\section{Fayet-Iliopoulos parameters}
\label{sect:fi}

For a number of years, it was thought that four-dimensional
$N=1$ supergravity could not admit moduli-independent Fayet-Iliopoulos
parameters, until the paper \cite{s0} observed a loophole:
Fayet-Iliopoulos parameters could be consistent so long as they
are quantized.  The papers \cite{ds,hs} observed that the
quantization condition is closely interrelated with the structure
of the Bagger-Witten line bundle.  Specifically, \cite{ds,hs} observed
that in a $G$ gauge theory, the (quantized) Fayet-Iliopoulos
parameter corresponds to a choice of $G$ action on the Bagger-Witten
line bundle.

Now, to be clear,
from the perspective of a low-energy observer, this matter is
somewhat moot:  a low-energy theory will only include a $U(1)$ if one
is working near a fixed point of its action on the moduli space, 
where the D-term vanishes.  Questions of nonzero Fayet-Iliopoulos
parameters are really questions about whether there can exist another
fixed point with a nonzero FI parameter, a question which has not to our
knowledge been addresses in the literature.  We will make a few observations
concerning these formal properties.

If the Bagger-Witten line bundle were trivial, then there would
be a natural `trivial' equivariant structure, which would correspond
to a vanishing Fayet-Iliopoulos parameter.  However, as observed elsewhere in this paper, that is not the case in any known examples.  When the Bagger-Witten
line bundle is nontrivial, as is the case in known examples,
there are two important subtleties to bear in mind:
\begin{itemize}
\item First, an action of the gauge group might not exist on the
Bagger-Witten line bundle, even if it is flat.
As a proof of concept, consider
the following example.  Let $E$ be an elliptic curve.
The line bundles on
$E$ that admit flat connections are parametrized by Pic$^0(E)$.
Consider the ${\mathbb Z}_2$ action on $E$ that sends $z \mapsto -z$.
Its induced action on Pic$^0(E)$ just sends a holomorphic line bundle
$L \mapsto L^{-1}$.  A line bundle in Pic$^0(E)$ will admit a
${\mathbb Z}_2$-equivariant structure if and only if it is fixed
under the action of the generator of ${\mathbb Z}_2$, in other
words if and only if $L \cong L^{-1}$, so that $L$ is a two-torsion
line bundle, of which there are precisely four.
Thus, although there is a continuous family of line
bundles with inequivalent flat connections on $E$, only four
of those admit a ${\mathbb Z}_2$-equivariant structure.
\item Second, there need be no natural `zero' equivariant
structure, hence no canonical way to make sense of a vanishing
Fayet-Iliopoulos parameter.
\end{itemize}
As we have seen explicitly in examples that the Bagger-Witten line
bundle is nontrivial, understanding the Fayet-Iliopoulos parameter in
four-dimensional supergravity is therefore rather nontrivial.
We hope to return to this matter in future work.

\section{Conclusions}

In this paper we have studied several aspects of the Bagger-Witten
line bundle, which arises naturally over moduli spaces of SCFTs and in
constructions of supergravity theories.  First, we proposed an intrinsic
definition of the Bagger-Witten line bundle over a moduli space
of Calabi-Yau's, as a line bundle of
covariantly constant spinors.  Second, we described a few more
concrete examples of Bagger-Witten line bundles.  This served two
purposes.  First, known examples are extremely rare in the literature.
Second, to construct the Bagger-Witten line bundle, one 
has to take a local trivially-acting ${\mathbb Z}_2$ quotient
(a ${\mathbb Z}_2$ gerbe), which 
reflects subtle dualities in the theory.  We worked through a few examples
of this process.  Next, we proposed a new criterion for existence of
UV completions of supergravity theories, namely that the Bagger-Witten
line bundle is a flat, nontrivial line bundle over the moduli space.
We also discussed some related mathematics conjectures.  Finally,
we discussed some subtleties in the application of these results to
target-space supergravity theories.

One matter that needs to be resolved is the form of anomaly computations.
Existing low-energy supergravity anomaly computations such as 
{\it e.g.} \cite{fk,efk} assume that the Bagger-Witten line bundle
is an honest line bundle over an ordinary moduli space, whereas
in fact we have seen that in typical examples the Bagger-Witten line
bundle is a `fractional' line bundle over a stack.  As discussed in
{\it e.g.} \cite{ajmos}, anomaly considerations could potentially
be more complicated, involving for example Chern class components over
associated inertia stacks that have no analogues for ordinary bundles
over smooth manifolds.

In this vein, the recent work \cite{Tachikawa:2017aux} applied the 
four-dimensional anomaly polynomial term
\begin{equation}
c_2(R) \left[ \frac{\omega}{2\pi} \right],
\end{equation}
for $\omega$ the K\"ahler form on the moduli space, to compute a formal
expression for the cohomology class of $\omega$ in moduli spaces of
four-dimensional $N=2$ SCFTs, by for example comparing to other terms
appearing in the four-dimensional anomaly polynomial via dimensional
reduction from six dimensions.  In two-dimensional theories, there
is an analogous term \cite{yujipriv} in anomaly polynomials of the form
\begin{equation}
c_1(R) \left[ \frac{\omega}{2\pi} \right].
\end{equation}
It would be interesting to repeat the arguments of \cite{Tachikawa:2017aux}
in a suitable context to formally derive expressions for the K\"ahler form
on the moduli space of two-dimensional SCFTs.

It would also be very interesting to apply these ideas to try to better
understand the K\"ahler potential on the moduli space.  Now, in a four
dimensional $N=1$ theory, this is not protected from any quantum corrections.
However, in the examples discussed here, it is constrained to be a 
modular-invariant function of complex structure moduli, which is a strong
constraint.  If that fact
were used in combination with some other constraints, it might be possible
to write a nearly exact expression for the K\"ahler potential, perhaps
something determined up to a few constants.

\section{Acknowledgements}

We would like to thank P.~Aspinwall, R.~Bryant, B.~Conrad, R.~Hain, 
S.~Katz, C.~Lazaroiu, I.~Melnikov, D.~Morrison, T.~Pantev,
and E.~Witten
for useful conversations.  
E.S. would like to thank Walter Parry for extensive
discussions, explanations, and computations 
of the groups Hom$(G, {\mathbb C}^{\times})$ for
various $G$ appearing in this paper,
and also C.~Hull and C.~Strickland-Constable
for discussions of duality group actions on fermions in
supergravity theories..
R.D. was partially supported by NSF grant DMS 1603526 and by
Simons Foundation grant number 390287.
E.S. was
partially supported by NSF grants PHY-1417410 and PHY-1720321.

\appendix

\section{Picard groups of moduli spaces of elliptic curves with level structures}
\label{app:picard}

This paper uses results for moduli spaces of
elliptic curves with two-torsion points and level two structures.
Although this material is well-known in the mathematics community,
it is much more obscure in the physics community, and in any event
many of the results we need are scattered across different 
sources.  To make this paper
self-contained,
we collect here a brief review of relevant results.

\subsection{Ordinary moduli spaces of elliptic curves}

First, recall that
the
fundamental domain for $SL(2,{\mathbb Z})$ is given by
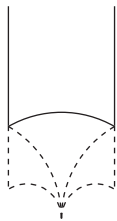
\begin{center}
\begin{picture}(40,80)
\Line(5,35)(5,80)
\Line(45,35)(45,80)
\CArc(25,0)(40,60,120)
\DashCArc(-15,0)(40,0,60){2}
\DashCArc(65,0)(40,120,180){2}
\DashCArc(12,0)(13,0,121){2}
\DashLine(5,12)(5,35){2}
\DashLine(45,12)(45,35){2}
\DashCArc(38,0)(13,59,180){2}
\end{picture}
\end{center}
Solid lines indicate the boundary for the usually-drawn fundamental domain
for $PSL(2,{\mathbb Z})$; dashed lines indicate the boundaries of a few
other nearby possible fundamental domains.  Curved edges arise from
circles of
radius 1 centered at integer points along the real axis, and circles of
radius $1/3$ centered at points $k/3$ along the real axis, for $k$ an
integer not divisible by 3.

There are two points on the fundamental domain with nontrivial stabilizers
(beyond the generic ${\mathbb Z}_2$).  One is the point $i$
(of $j$-invariant 1728), whose stabilizer is ${\mathbb Z}_4 \subset
SL(2,{\mathbb Z})$, which is generated by
\begin{equation}
S = \left[ \begin{array}{rr}
0 & -1 \\
1 & 0 \end{array} \right]. 
\end{equation}
Of these, $-I$ acts trivially on the upper-half-plane,
and so the point $i$ represents a local ${\mathbb Z}_2$ singularity
in the $PSL(2,{\mathbb Z})$ quotient.
The other point with nontrivial stabilizer 
is $\exp(2 \pi i/3) = -1/2 + i \sqrt{3}/2$.  This point has
stabilizer ${\mathbb Z}_6$, generated by $ST$ for $S$ as above and
\begin{equation}
T \: = \: \left[ \begin{array}{cc}
1 & 1 \\
0 & 1 \end{array} \right].
\end{equation}

In principle, we can read off the Picard group of the moduli space
from the two stabilizers above.  Briefly, using the fact that
$S^2 = (ST)^3 (=-I)$, we can identify the Picard group with
\begin{equation}
\frac{
{\mathbb Z}_4 \times {\mathbb Z}_6
}{
(g_1^2,1) \sim (1,g_2^3)
}
\end{equation}
(for $g_1$, $g_2$ the generators of the ${\mathbb Z}_4$, ${\mathbb Z}_6$
factors)
which can be shown to be isomorphic to ${\mathbb Z}_{12}$, with
generator $(g_1,g_2)$, corresponding to the Hodge line bundle.

Alternatively, 
the Picard group can be understood as the group of 
$SL(2,{\mathbb Z})$-equivariant line bundles on the upper half plane,
and since the upper half plane is simply connected, this is simply
\begin{equation}
{\rm Hom}(SL(2,{\mathbb Z}),{\mathbb C}^{\times}).
\end{equation}
This group is again ${\mathbb Z}_{12}$, corresponding to the abelianization
of $SL(2,{\mathbb Z})$, as determined by the generators $S$, $ST$
above.  (See for example \cite{litt} for a description of this approach and
\cite{kconrad} for a very readable discussion of the abelianization.)

In this language, the holonomy $\chi$ of the corresponding flat connection is
encoded as follows \cite{kconrad}:
\begin{equation}
\chi(S) \: = \: -i, \: \: \:
\chi(T) \: = \: \exp(2 \pi i/12) \: =\: -i \left( \frac{-1 + i \sqrt{3} }{
2} \right).
\end{equation}
In particular, for a compactification on an elliptic curve,
it seems that under the mirror to $B \mapsto B + 1$
around large-radius on the complex structure moduli space, corresponding
to the transformation $T$, we see that $\chi$ is nontrivial.

Extremely readable reviews of this case can be found in
\cite{fulton,hainrev,mumford,litt}.

In the next sections we will discuss the Picard groups of
${\cal M}_1(2) = [ {\mathfrak h}/\Gamma_1(2)]$ and
${\cal M}(2) = [ {\mathfrak h}/\Gamma(2)]$, where ${\mathfrak h}$ denotes
the upper half plane, and their relatives in
$PSL(2,{\mathbb Z})$ and $Mp(2,{\mathbb Z})$.  To that end, it may be
helpful to recall some basic relations between these subgroups of
$SL(2,{\mathbb Z})$.  First, $\Gamma(2)$ is a normal subgroup of
$SL(2,{\mathbb Z})$ with cokernel $S_3$:
\begin{equation}
1 \: \longrightarrow \: \Gamma(2) \: \longrightarrow \:
SL(2,{\mathbb Z}) \: \longrightarrow \: S_3 \: \longrightarrow \: 1.
\end{equation}
In addition, $\Gamma(2)$ is also a normal subgroup of $\Gamma_1(2)$:
\begin{equation}
1 \: \longrightarrow \: \Gamma(2) \: \longrightarrow \:
\Gamma_1(2) \: \longrightarrow \: {\mathbb Z}_2 \: \longrightarrow \: 1.
\end{equation}
Finally, $\Gamma_1(2)$ is a non-normal subgroup of $SL(2,{\mathbb Z})$
of index three.

\subsection{${\cal M}_1(2)$}

Define
\begin{equation}
\Gamma_1(m) \: \equiv \: \left\{ 
 \left[ \begin{array}{cc}
a & b \\
c & d
\end{array} \right] \: \equiv \:
\left[ \begin{array}{cc}
1 & * \\ 
0 & 1 \end{array} \right] \mbox{ mod } m \right\},
\end{equation}
a subgroup of $SL(2,{\mathbb Z})$.
(In other words, $a, d \equiv 1$ mod $m$ and $c \equiv 0$ mod $m$, but
$b$ is unconstrained.)

Define ${\cal M}_1(m) \equiv [ {\frak h}/\Gamma_1(m) ]$.
Then, ${\cal M}_1(m)$ can be interpreted as a moduli space of pairs $(E,p)$,
where $E$ is an elliptic curve and $p$ is a single $m$-torsion point
\cite{silverman}[appendix C.13].  This is because
for any $\gamma \in \Gamma_1(m)$,
\begin{equation}
\frac{1}{m} \: \mapsto \: \frac{1}{m (c \tau + d) } \: = \: \frac{1}{m} \: + \:
\frac{ (c/m) \tau + (d-1)/m }{c \tau+d} \: \sim \: \frac{1}{m},
\end{equation}
and so the $m$-torsion point $1/m$ is preserved.
In particular, ${\cal M}_1(2)$ is then a moduli space of elliptic curves with
a two-torsion point fixed.

A fundamental domain for $\Gamma_1(2) \subset SL(2,{\mathbb Z})$ is 
given by\footnote{
This was drawn with the assistance 
of Helena Verrill's fundamental domain drawer program, at
\href{https://wstein.org/Tables/fundomain/index2.html}{\tt https://wstein.org/Tables/fundomain/index2.html}.
} 
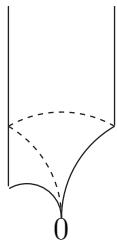
\begin{center}
\begin{picture}(40,80)
\Line(5,12)(5,80)
\Line(45,35)(45,80)
\DashCArc(25,0)(40,60,120){2}
\DashCArc(-15,0)(40,0,60){2}
\Text(25,0)[t]{$0$}
\CArc(65,0)(40,120,180)
\CArc(12,0)(13,0,121)
\end{picture}
\end{center}
where the left vertical edge has real part $-1/2$, the right vertical
edge has real part $+1/2$, and the curved edges arise from circles of
radius 1 centered at integer points along the real axis, and circles of
radius $1/3$ centered at points $k/3$ along the real axis, for $k$ an
integer not divisible by 3.
Solid lines are the boundaries of the the fundamental domain
of $\Gamma_1(2)$, 
whereas dashed lines indicate boundaries of fundamental
domains of $PSL(2,{\mathbb Z})$.  The vertex at the lower left is at
$-1/2 + i /(2 \sqrt{3})$, and the vertex at the upper right is at
$+1/2 + i \sqrt{3}/2$.  The fundamental domain above is essentially
one half of the fundamental domain for $\Gamma(2)$, as we shall see
shortly.

The lower left corner, at $-1/2 + i/(2 \sqrt{3})$, and the upper right corner,
at $+1/2 + i \sqrt{3}/2$, can be related by
\begin{equation}
\left[ \begin{array}{cc}
1 & 0 \\
2 & 1 \end{array} \right] \: \in \: \Gamma(2) \: \subset \: 
\Gamma_1(2),
\end{equation}
and so define the same point.  Similarly,
the point $-1/2 + i \sqrt{3}/2$, at the intersection of the
three fundamental domains for $SL(2,{\mathbb Z})$, is related to the
upper right point $+1/2 + i \sqrt{3}/2$ by the action of
\begin{equation}
\left[ \begin{array}{cc}
1 & 1\\
0 & 1 \end{array} \right] \: \in \: \Gamma_1(2).
\end{equation}
This particular matrix lies in $\Gamma_1(2)$ but not 
$\Gamma(2)$, and is
the essential reason why the fundamental domain for $\Gamma(2)$ looks like
two copies of the fundamental domain for $\Gamma_1(2)$.

The point $i$ is no longer a singular point, since the only elements
of $\langle S \rangle$ in $\Gamma_1(2)$ are the generic stabilizer $\pm I$.
However, the image of $i$ under $U = ST$ is the point
\begin{equation}
\alpha \: \equiv \: ST \cdot i \: = \: \frac{i}{2} - \frac{1}{2},
\end{equation}
lying along the left edge of the fundamental domain above,
and it has stabilizer ${\mathbb Z}_4 \subset \Gamma_1(2)$ generated by
\begin{equation}
K \: = \: U S U^{-1} \: = \: \left[ \begin{array}{rr}
-1 & -1 \\
2 & 1 \end{array} \right].
\end{equation}

In this case, as there is one point with stabilizer ${\mathbb Z}_4$,
for reasons similar to the ordinary moduli space, we see that 
Pic ${\cal M}_1(2) = {\mathbb Z}_4$, and it can be shown\footnote{
Strictly speaking, \cite{niles}[section 2] says that
Pic ${\cal M}_1(2)$ is a canonically split extension of ${\mathbb Z}_4$
by Pic $M_1(2)$, or more simply,
\begin{equation}
{\rm Pic}\, {\cal M}_1(2) \: = \:
{\rm Pic}\, M_1(2) \times {\mathbb Z}_4,
\end{equation}
where $M_1(2)$ is the coarse moduli space of ${\cal M}_1(2)$,
which is ${\mathbb P}^1$ minus two points, hence Pic $M_1(2)$ is trivial.
In passing, it may also be useful to the reader to note that
the $\Gamma_0(2)$ in that reference coincides with $\Gamma_1(2)$.
}
\cite{niles} that it is generated by the Hodge line bundle.
Note in passing that this means the Hodge line bundle is fractional,
a line bundle on the gerbe which is not a pullback from the underlying space.

We now turn to Picard groups and Hodge line bundles on
$[ {\mathfrak h}/\tilde{\Gamma}_1(2) ]$ and
$[ {\mathfrak h}/p^* \Gamma_1(2)]$, for $\tilde{\Gamma}_1(2) \subset
PSL(2,{\mathbb Z})$ and $p: Mp(2,{\mathbb Z}) \rightarrow SL(2,{\mathbb Z})$.

First, note that there are projection maps
\begin{equation}
[ {\mathfrak h}/p^* \Gamma_1(2)] \: \longrightarrow \:
[ {\mathfrak h}/ \Gamma_1(2) ] \: \longrightarrow \:
[ {\mathfrak h}/\tilde{\Gamma}_1(2) ] \: \longrightarrow \:
{\mathfrak h} / \tilde{\Gamma}_1(2).
\label{eq:related-stacks}
\end{equation}
We construct the last space, the variety ${\mathfrak h}/\tilde{\Gamma}_1(2)$,
as follows.  
Begin with the variety ${\mathfrak h}/\Gamma(2)$, which is
\begin{equation}
{\cal M}_{0,4} \: = \: {\mathbb P}^1 \: - \: \mbox{(3 points)},
\end{equation}
where the three points are $\{ 0, 1, \infty\}$.
We can now quotient this space by a transposition of order two, defined by
an involution of $S_3$, to get the variety
${\mathfrak h}/\tilde{\Gamma}_1(2)$.  We can describe possible actions 
on ${\mathbb P}^1$ explicitly as follows:
\begin{enumerate}
\item $\lambda \mapsto \lambda$, the identity,
\item $\lambda \mapsto 1/(1-\lambda)$, which sends $0 \rightarrow 1 
\rightarrow \infty \rightarrow 0$ and has order three,
\item $\lambda \mapsto \lambda - 1/\lambda$, which sends
$0 \rightarrow \infty \rightarrow 1 \rightarrow 0$ and has order three,
\item $\lambda \mapsto + 1/\lambda$, which sends $0 \leftrightarrow \infty$,
has order two,
and leaves the points $\pm 1$ fixed,
\item $\lambda \mapsto \lambda / (\lambda - 1)$, which sends
$1 \leftrightarrow \infty$, has order two, and leaves fixed $0, 2$,
\item $\lambda \mapsto 1-\lambda$, which sends $0 \leftrightarrow 1$,
has order two,
and leaves fixed $\infty$, $1/2$.
\end{enumerate}
To be definite, consider the quotient by the last involution.
On the quotient, which is the variety ${\mathfrak h}/
\tilde{\Gamma}_1(2)$, there are now two excluded points
(namely $\infty$ and $0=1$), and one marked point ($1/2$) of nontrivial
stabilizer.  Denote that marked point by $y$.

The stack $[ {\mathfrak h}/ \tilde{\Gamma}_1(2)]$ is isomorphic
to the variety ${\mathfrak h}/ \tilde{\Gamma}_1(2)$ away from the
marked point $y$.  Over $y$, the stack $[ {\mathfrak h}/\tilde{\Gamma}_1(2)]$
has stabilizer ${\mathbb Z}_2$.  Denote this stabilizer $Z^p$.

The stack $[ {\mathfrak h}/\Gamma_1(2)]$ is a ${\mathbb Z}_2$ gerbe
over $[ {\mathfrak h}/\tilde{\Gamma}_1(2)]$, and over $y$ has stabilizer
${\mathbb Z}_4$.  Denote this stabilizer $Z$.

The stack $[ {\mathfrak h}/p^* \Gamma_1(2)]$ is a ${\mathbb Z}_4$ gerbe
over $[ {\mathfrak h}/\tilde{\Gamma}_1(2)]$, and we will see below
that
over $y$ it has
stabilizer ${\mathbb Z}_8$.
Let $Z^m$ denote the stabilizer over $y$,
then from the sequence~(\ref{eq:related-stacks}) we have the
relation between stabilizers
\begin{equation}
Z^m \: \longrightarrow \: Z={\mathbb Z}_4 \: \longrightarrow \:
Z^p = {\mathbb Z}_2.
\end{equation}
We know that $Z^m \rightarrow Z={\mathbb Z}_4$ is surjective, with kernel
equal to the kernel of $p: Mp(2,{\mathbb Z}) \rightarrow SL(2,{\mathbb Z})$,
namely ${\mathbb Z}_2$, and quotient ${\mathbb Z}_4$, hence we know $Z^m$
is given by an extension
\begin{equation}
1 \: \longrightarrow \: {\mathbb Z}_2 \: \longrightarrow \:
Z^m \: \longrightarrow \: Z={\mathbb Z}_4 \: \longrightarrow \: 1.
\label{eq:stab-y-mp}
\end{equation}
It remains to determine whether the extension $Z$ is ${\mathbb Z}_2 \times
{\mathbb Z}_4$ or ${\mathbb Z}_8$.  If we pullback along the center of 
$SL(2,{\mathbb Z})$, which preserves the kernel ${\mathbb Z}_2$ above,
this becomes the nontrivial extension
\begin{equation}
1 \: \longrightarrow \: {\mathbb Z}_2 \: \longrightarrow \: {\mathbb Z}_4
\: \longrightarrow \: {\mathbb Z}_2 \: \longrightarrow \: 1,
\end{equation}
hence the extension~(\ref{eq:stab-y-mp}) 
must be nontrivial, and so $Z^m = {\mathbb Z}_8$.
Finally, as these stabilizers live inside gerbe structures, they define
twists by line bundles (on the gerbes), and so we see in each case
that the Picard group matches the stabilizer at $y$, as summarized
in table~\ref{table:summ}.

Since the Hodge line bundle is acted upon by the center of 
$SL(2,{\mathbb Z})$, it does not exist over $[ {\mathfrak h} / 
\tilde{\Gamma}_1(2)]$, and its pullback to
$[ {\mathfrak h}/p^* \Gamma_1(2)]$ is the square of the generator of the
Picard group there.

\subsection{${\cal M}(2)$}

Define
${\cal M}(m) = [ {\frak h}/\Gamma(m) ]$,
where $\Gamma(m)$ is the (``principal congruence'') subgroup of
$SL(2,{\mathbb Z})$ consisting of matrices equivalent to the identity
mod $m$:
\begin{equation}
\Gamma(m) \: \equiv \: \left\{
\left. \left[ \begin{array}{cc}
a & b \\
c & d
\end{array} \right] \right| a \equiv 1 \mbox{ mod } m, d \equiv 1 \mbox{ mod }
m, b \equiv 0 \mbox{ mod }m, c \equiv 0 \mbox{ mod } m \right\}.
\end{equation}
(The notation $\Gamma(m)$ or $SL(2,{\mathbb Z})[m]$ is sometimes used
instead; however, $\Gamma(m)$ is also sometimes used to refer to the
analogous subgroup of $PSL(2,{\mathbb Z})$, so to try to remove ambiguity,
we will use $\Gamma(m)$ or $SL(2,{\mathbb Z})[m]$ to denote
this subgroup of $SL(2,{\mathbb Z})$.)

The moduli space ${\cal M}(m)$ defined above is the moduli stack
of elliptic curves with a level $m$ structure \cite{hainrev}[section 4.2].
The idea is that $\Gamma(m)$ preserves the level $m$ structure.

To gain a bit of intuition, note that for $m=2$, $\Gamma(2)$ preserves
a choice of spin structure on $T^2$.  If we let periodicities around two cycles
be denoted $(-)^m$, $(-)^n$, then under the action of an element of
$\Gamma(2)$ above,
\begin{equation}
\left[ \begin{array}{c} (-)^m \\ (-)^n \end{array} \right] \: \mapsto
\: \left[ \begin{array}{c} (-)^{am+bn} \\ (-)^{cm_dn} \end{array} \right]
\: = \: 
\left[ \begin{array}{c} (-)^m \\ (-)^n \end{array} \right],
\end{equation}
where the last equality follows from the fact that for an element
\begin{equation}
\left[ \begin{array}{cc}
a & b \\ c & d \end{array} \right] \: \in \: \Gamma(2),
\end{equation}
$a$ and $d$ are odd, while $b$ and $c$ are even.

We can also interpret ${\cal M}(m)$ as a moduli space of
elliptic curves with a basis of $m$-torsion points
\cite{silverman}[appendix C.13].  The basic point is that
$\gamma \in \Gamma(m)$ will preserve the $m$-torsion points
including $1/m$ and $\tau/m$.
For example, for such $\gamma$,
\begin{eqnarray*}
\frac{1}{m} \: \mapsto \: \frac{1}{m (c \tau + d) } & = & \frac{1}{m} \: + \:
\frac{ (c/m) \tau + (d-1)/m }{c \tau+d} \: \sim \: \frac{1}{m}, \\
\frac{\tau}{m} \: \mapsto \: \frac{\tau}{m (c \tau + d)} & = &
\frac{1}{m} \frac{a \tau + b}{c \tau + d} \: + \:
\frac{d-1}{m} \frac{a \tau + b}{c \tau + d} \: - \: \frac{b}{m}
\: \sim \: \frac{\tau}{m},
\end{eqnarray*}
where we have used the fact that $b, c \equiv 0$ mod $m$
and $d-1 \equiv 0$ mod $m$.  Similarly, one can show that $k/m$ is invariant
under $\gamma$ for integer $k$, as are other $m$-torsion points.
For this reason, a level $m$ structure is
equivalent to a basis $\{ 1/m, \cdots,
\tau/m \}$ for the $m$-torsion points.

In passing, the moduli space ${\cal M}(m)$ has automorphisms given by
$SL(2,{\mathbb Z}) / \Gamma(m) \cong SL(2,{\mathbb Z}_m)$.
For example, for $m=2$, these automorphisms form $S_3$, and permute the
three nontrivial two-torsion points of a fixed elliptic curve.

A fundamental domain for $\Gamma(2)$ is illustrated below
\cite{katok}[section 5.5, fig. 28]:
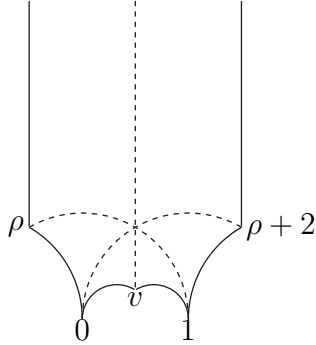
\begin{center}
\begin{picture}(90,120)
\Line(5,35)(5,120)
\DashLine(45,12)(45,120){2}
\Line(85,35)(85,120)
\Text(3,35)[r]{$\rho$}
\Text(87,35)[l]{$\rho+2$}
\Text(25,0)[t]{$0$}
\Text(65,0)[t]{$1$}
\CArc(105,0)(40,120,180)
\CArc(-15,0)(40,0,60)
\DashCArc(25,0)(40,0,120){2}
\DashCArc(65,0)(40,60,180){2}
\CArc(38,0)(13,59,180)
\CArc(52,0)(13,0,121)
\Text(45,10)[t]{$v$}
\end{picture}
\end{center}
Solid lines indicate boundaries of the fundamental domain of
$\Gamma(2)$.
Dashed lines define boundaries between fundamental domains of
$PSL(2,{\mathbb Z})$.  The straight vertical boundaries lie along lines
of real part $-1/2$, $+3/2$.
The curved boundaries lie along circles of
radius 1 centered at integer points along the real axis, and also
circles of radius $1/3$ centered at points $k/3$ along the real
axis, for $k$ an integer not divisible by 3.
The vertices $\rho = -1/2+i \sqrt{3}/2$,
$\rho+2$, $v =  1/2 + i/(2
\sqrt{3})$
are related by $\Gamma(2)$, and so define the same point.
Clearly, the fundamental domain for $\Gamma(2)$ contains six copies of
fundamental domains of $SL(2,{\mathbb Z})$.
The one-point compactification of this domain is a sphere with
three cusps, at $0$, $1$, and $\rho = \rho+2 = v$.

The reader should note that the fundamental domain above for
$\Gamma(2)$ is precisely two copies of the fundamental domain
for $\Gamma_1(2)$ given previously.  The difference comes down to
$T \in \Gamma_1(2)$ which is not also in $\Gamma(2)$.

The reader should also note that the point $\alpha$, which lies on the left
edge of the fundamental domain previously discussed for $\Gamma_1(2)$
but is in the middle of the fundamental region above for $\Gamma(2)$,
is no longer singular:  the only elements of $\langle K \rangle$ which lie
in $\Gamma(2)$ are $\pm I$, the stabilizer of generic points.
We can also see this in another way.  Recall that the fundamental
domain of $SL(2,{\mathbb Z})$ had points with stabilizers $\langle S \rangle$
and $\langle S T \rangle$.  It is straightforward to check that most
elements of $\langle S \rangle$ and $\langle S T \rangle$ are not in
$\Gamma(2)$, with the exception of the ${\mathbb Z}_2$ center of 
$SL(2,{\mathbb Z})$.  As a result, at those potentially problematic
points, the only stabilizer is the same as at every other point,
the generic stabilizer defining the ${\mathbb Z}_2$ gerbe structure.
The other elements of $\langle S \rangle$ and $\langle S T \rangle$
simply move between different copies of the fundamental domain of
$SL(2,{\mathbb Z})$, within the fundamental domain of 
$\Gamma(2)$.

This is a ${\mathbb Z}_2$ gerbe over the space
$M_{0,4} = {\mathbb C}-\{0,1\}$.  More formally, we can describe elliptic
curves as cubics in ${\mathbb P}^2 = {\rm Proj}\, {\mathbb C}[x,y,z]$
of the form
\begin{equation}
y^2 z \: = \:  x (x-z) (x-\lambda z),
\end{equation}
where $\lambda$ parametrizes the family.  The two-torsion points are then
\cite{silvermantate}[chapter II.1]
\begin{equation}
(x,y,z) \: = \: (0,0,1), \:
(1,0,1), \: (\lambda,0,1), \: (0,1,0),
\end{equation}
where the point with $y=1$ corresponds to the origin and the
remaining points, of $y=0$, correspond to the nonzero two-torsion points.
${\cal M}(2)$ is a moduli stack of elliptic curves with a basis of 
two-torsion points, and at $\lambda = 0, 1, \infty$, some of the
two-torsion points collide, so we exclude those points.
Hence, ${\cal M}(2)$ is a gerbe over 
\begin{equation}
M_{0,4} \: = \: {\mathbb P}^1 - \{ 0, 1, \infty \} \: = \:
{\mathbb C} - \{0, 1\}.
\end{equation}
Now, $H^2(M_{0,4}, {\mathbb Z}_k) = 0$, so any ${\mathbb Z}_2$ gerbe
over $M_{0,4}$ is trivial, hence ${\cal M}(2)$ is the trivial
${\mathbb Z}_2$ gerbe over $M_{0,4}$.

Now we can compute the Picard group.  Since ${\cal M}(2)$ is a trivial
${\mathbb Z}_2$ gerbe on $M_{0,4}$,
\begin{equation}
{\rm Pic}\, {\cal M}(2) \: = \: {\mathbb Z}_2 \times
{\rm Pic}\, M_{0,4}.
\end{equation}
However, Pic $M_{0,4} = 0$, so we see that Pic ${\cal M}(2) = {\mathbb Z}_2$.

The Hodge line bundle on ${\cal M}(2)$ is the pullback\footnote{
One way to construct a map from Pic ${\cal M}$ to Pic ${\cal M}(2)$ is
as follows.  First, since $\Gamma(2)$ is a normal subgroup of 
$SL(2,{\mathbb Z})$ with cokernel $S_3$,
\begin{equation}
1 \: \longrightarrow \: \Gamma(2) \: \longrightarrow \: SL(2,{\mathbb Z})
\: \longrightarrow \: S_3 \: \longrightarrow \: 1,
\end{equation}
Pic ${\cal M}$ is isomorphic to $S_3$-equivariant line bundles on
${\cal M}(2)$.  More generally, given
\begin{equation}
1 \: \longrightarrow \: K \: \longrightarrow \: G \: \longrightarrow \: H 
\: \longrightarrow \: 1,
\end{equation}
line bundles on $[X/G]$ are $H$-equivariant line bundles on $[X/K]$.
Given an $S_3$-equivariant line bundle on ${\cal M}(2)$, we can
forget the $S_3$-equivariant structure to map to Pic ${\cal M}(2)$.
In other words,
\begin{equation}
{\rm Pic}\, {\cal M} \: \longrightarrow \:
{\rm Pic}^{S_3}\, {\cal M}(2) \: \longrightarrow \: {\rm Pic}\, {\cal M}(2).
\end{equation}
} of the Hodge line
bundle on ${\cal M}$, and is necessarily nontrivial judging from the
nontrivial action of the center of $SL(2,{\mathbb Z})$.  (In passing,
this means the pullback from Pic ${\cal M}$ to Pic ${\cal M}(2)$ is surjective.)
Thus, we see that
the Picard group of ${\cal M}(2)$ is generated by the Hodge line bundle,
which is fractional (a line bundle on the gerbe which is not a pullback
from the underlying space). 

In passing, note that if we define $\tilde{\Gamma}(2)$ to be the analogous
subgroup of $PSL(2,{\mathbb Z})$, then trivially
$[ {\mathfrak h} / \tilde{\Gamma}(2) ] = M_{0,4}$, hence
the Picard group of $[ {\mathfrak h} / \tilde{\Gamma}(2) ]$ is trivial.

Similarly, for $p$ the projection 
$Mp(2,{\mathbb Z}) \rightarrow SL(2,{\mathbb Z})$, note
$[ {\mathfrak h} / p^* \Gamma(2) ]$ is a ${\mathbb Z}_2$ gerbe over
$[ {\mathfrak h} / \Gamma(2) ]$ and (since $Mp(2,{\mathbb Z})$ is a 
${\mathbb Z}_4$ extension of $PSL(2,{\mathbb Z})$ a ${\mathbb Z}_4$ gerbe
over $[ {\mathfrak h}/\tilde{\Gamma}(2)] = M_{0,4}$.  Since there are
no nontrivial root gerbes over $M_{0,4}$, it must be the trivial gerbe,
hence
\begin{equation}
{\rm Pic}\, [ {\mathfrak h}/ p^* \Gamma(2) ] \: = \:
{\mathbb Z}_4 \times {\rm Pic}\, M_{0,4}.
\end{equation}
Since the Picard group of $M_{0,4}$ is trivial, we then have that the
Picard group of $[ {\mathfrak h}/ p^* \Gamma(2) ]$ is 
${\mathbb Z}_4$.

\subsection{Flat connections on moduli spaces}

Although it is not directly relevant to the analyses of the rest of the paper,
it seems appropriate to also list related results
concerning flat $U(1)$ connections over the moduli spaces of
elliptic curves with level structures, which is what we shall do in
this section.

The group $\tilde{\Gamma}(2) \subset PSL(2,{\mathbb Z})$ is a free group
on 2 generators.  (In fact, it can be identified with
$\pi_1( {\mathbb C} -\{0,1\})$.)
We can take the generators to be
\begin{equation}
\left[ \begin{array}{cc}
1 & 2 \\ 0 & 1 \end{array} \right], \: \: \:
\left[ \begin{array}{cc}
1 & 0 \\ 2 & 1 \end{array} \right].
\end{equation}
As a result,
\begin{equation}
{\rm Hom}(\tilde{\Gamma}(2), {\mathbb C}^{\times}) \: = \:
{\mathbb C}^{\times} \times
{\mathbb C}^{\times}.
\end{equation}
In $SL(2,{\mathbb Z})$, $\Gamma(2)$ contains $-I$, generating ${\mathbb Z}_2$,
as well as the two matrices above, which generate a subgroup of
$\Gamma(2)$ isomorphic to $\tilde{\Gamma}(2)$,
hence
$\Gamma(2) \cong {\mathbb Z}_2 \times \tilde{\Gamma}(2)$,
and
\begin{equation}
{\rm Hom}( \Gamma(2), {\mathbb C}^{\times}) \: = \:
{\mathbb Z}_2 \times 
{\rm Hom}( \tilde{\Gamma}(2), {\mathbb C}^{\times}) \: = \:
{\mathbb Z}_2 \times {\mathbb C}^{\times} \times
{\mathbb C}^{\times}.
\end{equation}

The group $\Gamma_1(2) \subset SL(2,{\mathbb Z})$
can be defined either as the group of matrices
\begin{equation}
\left[ \begin{array}{cc}
a & b \\
c & d \end{array} \right]
\end{equation}
such that
\begin{enumerate}
\item $a \cong 1 \mbox{ mod }2$, $d \cong 1 \mbox{ mod }2$,
$c \cong 0 \mbox{ mod }2$, which is the specialization of
the definition for $\Gamma_1(N)$ for general $N$, or equivalently for the
case $N=2$,
\item $c \cong 0 \mbox{ mod }2$.
\end{enumerate}
We can see that the second implies the first, for $N=2$, as follows:
if $c$ is even, then in order for the determinant to be 1, neither
diagonal entry can be even.  The result follows.  Technically, the second
case is sometimes denoted $\Gamma_0(2)$, hence this means that
$\Gamma_1(2) = \Gamma_0(2)$.

Let $\tilde{\Gamma}_1(2) \subset PSL(2,{\mathbb Z})$ denote the
image of $\Gamma_1(2)$.
The group $\tilde{\Gamma}_1(2)$ is also a free product, but of
${\mathbb Z}_2$ and ${\mathbb Z}$ instead of two copies of ${\mathbb Z}$.
The generators are the images in $PSL(2,{\mathbb Z})$ of
\begin{equation}  \label{eq:gamma-1-2:generators}
\left[ \begin{array}{cc}
1 & -1 \\
2 & -1 \end{array} \right], \: \: \:
\left[ \begin{array}{cc}
1 & 1 \\
0 & 1 \end{array} \right].
\end{equation}
The first matrix has order 4 in $SL(2,{\mathbb Z})$, and its image in
$PSL(2,{\mathbb Z})$ has order 2, whereas the second matrix has infinite
order, hence
\begin{eqnarray*}
{\rm Hom}( \tilde{\Gamma}_1(2), {\mathbb C}^{\times} ) & = &
{\mathbb Z}_2 \times {\mathbb C}^{\times}.
\end{eqnarray*}

In this case, $\Gamma_1(2)$ does not contain
a subgroup isomorphic to $\tilde{\Gamma}_1(2)$.
The matrices~(\ref{eq:gamma-1-2:generators}) generate
$\Gamma_1(2)$.
More formally, $\Gamma_1(2)$ is the free product with
amalgamation of ${\mathbb Z}_4$, generated by the first matrix,
and ${\mathbb Z}_2 \times {\mathbb Z}$, generated by $\pm 1$ times the
second matrix, with amalgamation along the common subgroups of order 2.
Hence,
\begin{equation}
{\rm Hom}( \Gamma_1(2), {\mathbb C}^* ) \: = \:
{\mathbb Z}_4 \times {\mathbb C}^{\times}.
\end{equation}

Since $PSL(2,{\mathbb Z})$ is the free product
of ${\mathbb Z}_2$ and ${\mathbb Z}_3$, its abelianization is
${\mathbb Z}_6$ and
\begin{equation}
{\rm Hom}( PSL(2,{\mathbb Z}), {\mathbb C}^{\times}) \: = \:
{\mathbb Z}_6.
\end{equation}
Similarly, the abelianization of $SL(2,{\mathbb Z})$ is
${\mathbb Z}_{12}$ and
\begin{equation}
{\rm Hom}( SL(2,{\mathbb Z}), {\mathbb C}^{\times}) \: = \:
{\mathbb Z}_{12}.
\end{equation}

Let $p: Mp(2,{\mathbb Z}) \rightarrow SL(2,{\mathbb Z})$ be projection.
Recall $Mp(2,{\mathbb Z})$ has a unique nontrivial element of order 2,
and $p^* -1$ is the set of 2 elements of order 4 which are central in
$Mp(2,{\mathbb Z})$.

As a result, $p^* \Gamma(2)$ is the direct product of ${\mathbb Z}_4$
and a free group of rank 2, hence
\begin{equation}
{\rm Hom}( p^* \Gamma(2), {\mathbb C}^{\times}) \: = \:
{\mathbb Z}_4 \times {\mathbb C}^{\times} \times
{\mathbb C}^{\times}.
\end{equation}
Similarly,
\begin{equation}
{\rm Hom}( p^* \Gamma_1(2), {\mathbb C}^{\times}) \: = \:
{\mathbb Z}_8 \times {\mathbb C}^{\times}.
\end{equation}

As a technical aside, note that the abelianization of a group $G$ is
the set of cocharacters of Hom$(G, {\mathbb C}^{\times})$, not the
Hom group itself.  For example,
\begin{equation}
{\rm Hom}( \Gamma(2), {\mathbb C}^{\times}) \: = \:
{\mathbb Z}_2 \times {\mathbb C}^{\times} \times {\mathbb C}^{\times},
\end{equation}
but the abelianization of $\Gamma(2)$ is ${\mathbb Z}_2 \times {\mathbb Z}
\times {\mathbb Z}$.  (In particular, the abelianization of a discrete
group cannot contain a ${\mathbb C}^{\times}$.)

\subsection{Summary}

In table~\ref{table:summ} we summarize the results of this appendix,
on Picard groups, Hodge line bundles, and flat connections on
stacks of the form $[ {\mathfrak h}/G]$, for ${\mathfrak h}$ the
upper half plane.

\begin{table}
\begin{center}
\begin{tabular}{c|c|c|c}
Group $G$ & Pic $[{\mathfrak h}/G]$ & Hodge & Hom$(G,{\mathbb C}^{\times})$ \\ \hline
$PSL(2,{\mathbb Z})$ & ${\mathbb Z}_6$  & --- & ${\mathbb Z}_6$ \\
$SL(2,{\mathbb Z})$ & ${\mathbb Z}_{12}$  & $g$ & ${\mathbb Z}_{12}$ \\
$Mp(2,{\mathbb Z})$ & ${\mathbb Z}_{24}$ & $g^2$ & ${\mathbb Z}_{24}$ \\ \hline
$\tilde{\Gamma}(2) \subset PSL(2,{\mathbb Z})$ &
$1$ & --- &
${\mathbb C}^{\times} \times {\mathbb C}^{\times}$ \\
$\Gamma(2) \subset SL(2,{\mathbb Z})$ & ${\mathbb Z}_2$ & $g$ & 
${\mathbb Z}_2 \times
{\mathbb C}^{\times} \times {\mathbb C}^{\times}$ \\
$p^* \Gamma(2) \subset Mp(2,{\mathbb Z})$ &  ${\mathbb Z}_4$ & $g^2$ &
${\mathbb Z}_4 \times {\mathbb C}^{\times} \times
{\mathbb C}^{\times}$ \\ \hline
$\tilde{\Gamma}_1(2) \subset PSL(2,{\mathbb Z})$ & ${\mathbb Z}_2$ & --- &
${\mathbb Z}_2 \times {\mathbb C}^{\times}$ \\
$\Gamma_1(2) \subset SL(2,{\mathbb Z})$ & ${\mathbb Z}_4$ & $g$ &
${\mathbb Z}_4 \times {\mathbb C}^{\times}$ \\
$p^* \Gamma_1(2) \subset Mp(2,{\mathbb Z})$ & ${\mathbb Z}_8$ & $g^2$ &
${\mathbb Z}_8 \times {\mathbb C}^{\times}$
\end{tabular}
\caption{Listed here are Picard groups, Hodge line bundles, and Hom's
for the examples described in this appendix.  In the case of Hodge line
bundles, $g$ indicates the generator of the Picard group, and --- indicates
that the Hodge line bundle is not canonically defined. \label{table:summ}}
\end{center}
\end{table}

Note that for each of $[ {\mathfrak h} / SL(2,{\mathbb Z}) ]$,
$[ {\mathfrak h}/ \Gamma(2) ]$, and $[ {\mathfrak h} / \Gamma_1(2) ]$,
the Hodge line bundle generates the Picard group.  As a result, in each
case, to construct a square root of the Hodge line bundle, one must
replace the original quotient by a quotient by a ${\mathbb Z}_2$
extension.  In particular, in
$[ {\mathfrak h} / Mp(2,{\mathbb Z}) ]$,
$[ {\mathfrak h} / p^* \Gamma(2) ]$,
$[ {\mathfrak h} / p^* \Gamma_1(2) ]$,
the Hodge line bundle is the square of the generator of the Picard
group, so we can identify the Bagger-Witten line bundle with the
generator itself.

It is also worth noting that in each of the cases above,
the Hodge line bundle only exists over a gerby quotient --
the Hodge line bundle does not exist over the
effectively-acting quotients
$ {\mathfrak h} / PSL(2,{\mathbb Z})$,
${\mathfrak h} / \tilde{\Gamma}(2)$,
or ${\mathfrak h} / \tilde{\Gamma}_1(2)$.


\begin{thebibliography}{199}

\addcontentsline{toc}{section}{References}

\bibitem{bw} E. Witten, J. Bagger, ``Quantization of Newton's constant
in certain supergravity theories,'' Phys. Lett. {\bf B115} (1982) 
202-206.

\bibitem{ps} V. Periwal, A. Strominger, ``K\"ahler geometry of the space
of $N=2$ superconformal field theories,'' Phys. Lett. {\bf B235} (1990)
261-267.

\bibitem{distsigma} J. Distler, ``Notes on N=2 sigma models,''
pp. 234-256 in {\it String theory and quantum gravity '92 (Proceedings,
Trieste 1992)}, {\tt hep-th/9212062}.

\bibitem{Bershadsky:1993cx}
  M.~Bershadsky, S.~Cecotti, H.~Ooguri and C.~Vafa,
  ``Kodaira-Spencer theory of gravity and exact results for quantum string amplitudes,''
  Commun.\ Math.\ Phys.\  {\bf 165} (1994) 311-428, 
  {\tt hep-th/9309140}.

\bibitem{mebw} W. Gu, E. Sharpe, ``Bagger-Witten line bundles on moduli
spaces of elliptic curves,'' Int. J. Mod. Phys. {\bf A31} (2016)
1650188, {\tt arXiv:1606.07078}.

\bibitem{ds17} R. Donagi, M. Macerato,
E. Sharpe, ``On the global moduli of Calabi-Yau
threefolds,'' {\tt arXiv:1707.05322}.

\bibitem{Gomis:2015yaa}
  J.~Gomis, P.~S.~Hsin, Z.~Komargodski, A.~Schwimmer, N.~Seiberg and S.~Theisen,
  ``Anomalies, conformal manifolds, and spheres,''
  JHEP {\bf 1603} (2016) 022,
  {\tt arXiv:1509.08511}.

\bibitem{meta} T. Pantev, E. Sharpe, ``Duality group actions
on fermions,'' JHEP {\bf 1611} (2016) 171,
{\tt arXiv:1609.00011}.

\bibitem{Donagi:2017vwh}
  R.~Donagi and D.~R.~Morrison,
  ``Conformal field theories and compact curves in moduli spaces,''
JHEP {\bf 1805} (2018) 021,
  {\tt arXiv:1709.05355}.

\bibitem{Tachikawa:2017aux}
  Y.~Tachikawa,
  ``Anomalies involving the space of couplings and the Zamolodchikov metric,''
JHEP {\bf 1712} (2017) 140,
  {\tt arXiv:1710.03934}.

\bibitem{Hull:2007zu}
  C.~M.~Hull,
  ``Generalised geometry for M-theory,''
  JHEP {\bf 0707} (2007) 079,
  {\tt hep-th/0701203}.

\bibitem{Keurentjes:2003hc}
  A.~Keurentjes,
  ``U duality (sub)groups and their topology,''
  Class.\ Quant.\ Grav.\  {\bf 21} (2004) S1367-1374,
  {\tt hep-th/0312134}.
 
\bibitem{ds} J. Distler, E. Sharpe, ``Quantization of Fayet-Iliopoulos
parameters in supergravity,'' Phys. Rev. {\bf D83} (2011)
085010,
{\tt arXiv:1008.0419}.
  
\bibitem{Tateishi:2018rnq}
  A.~D.~Tateishi,
  ``Quantum correction from super-Weyl transformation in supergravity,''
  {\tt arXiv:1806.07622}.
  
\bibitem{wb} J. Wess, J. Bagger, {\it Supersymmetry and supergravity},
Princeton University Press, Princeton, NJ, second edition, 1992. 
  
\bibitem{hs} S. Hellerman, E. Sharpe, ``Sums over topological sectors
and quantization of Fayet-Iliopoulos parameters,'' Adv. Theor.
Math. Phys. {\bf 15} (2011) 1141-1199,
{\tt arXiv:1012.5999}.
  
\bibitem{ajmos} L. Anderson, B. Jia, R. Manion, B. Ovrut, E. Sharpe,
``General aspects of heterotic string compactifications on stacks
and gerbes,'' Adv. Theor. Math. Phys. {\bf 19} (2015) 531-611,
{\tt arXiv:1307.2269}. 
 

\bibitem{Lerche:1989uy} 
  W.~Lerche, C.~Vafa and N.~P.~Warner, 
  ``Chiral rings in N=2 superconformal theories,''
  Nucl.\ Phys.\ B {\bf 324} (1989) 427-474.



 
\bibitem{hainrev} R. Hain, ``Lectures on moduli spaces of
elliptic curves,'' pp. 95-166 in {\it Transformation groups and
moduli spaces of curves}, ed. L. Ji, S.-T. Yau, Advanced Lectures in
Mathematics, vol. 16, International Press, 2011,
{\tt arXiv:0812.1803}.
  
\bibitem{lm} H. Blaine Lawson, Jr., M.-L. Michelsohn,
{\it Spin geometry}, Princeton University Press, Princeton, NJ, 1989.

\bibitem{Sharpe:2013bwa}
  E.~Sharpe,
  ``A few Ricci-flat stacks as phases of exotic GLSM's,''
  Phys.\ Lett.\ B {\bf 726} (2013) 390-395,
  {\tt arXiv:1306.5440}.

\bibitem{Gauntlett:2003cy}
  J.~P.~Gauntlett, D.~Martelli and D.~Waldram,
  ``Superstrings with intrinsic torsion,''
  Phys.\ Rev.\ D {\bf 69} (2004) 086002,
  {\tt hep-th/0302158}.

\bibitem{gsw} M. Green, J. Schwarz, E. Witten, {\it Superstring theory,
volume 2: loop amplitudes, anomalies and phenomenology}, Cambridge University
Press, New York, 1987.

\bibitem{Hebecker:2017lxm}
  A.~Hebecker, P.~Henkenjohann and L.~T.~Witkowski,
  ``Flat monodromies and a moduli space size conjecture,''
  JHEP {\bf 1712} (2017) 033,
  {\tt arXiv:1708.06761}.

\bibitem{Ooguri:2006in}
  H.~Ooguri and C.~Vafa,
  ``On the geometry of the string landscape and the swampland,''
  Nucl.\ Phys.\ B {\bf 766} (2007) 21-33,
  {\tt hep-th/0605264}.

\bibitem{Brennan:2017rbf}
  T.~D.~Brennan, F.~Carta and C.~Vafa,
  ``The string landscape, the swampland, and the missing corner,''
PoS TASI {\bf 2017} (2017) 015,
  {\tt arXiv:1711.00864}.

\bibitem{ArkaniHamed:2006dz}
  N.~Arkani-Hamed, L.~Motl, A.~Nicolis and C.~Vafa,
  ``The string landscape, black holes and gravity as the weakest force,''
  JHEP {\bf 0706} (2007) 060,
  {\tt hep-th/0601001}.



\bibitem{ogg} A. Ogg,
``Automorphismes de courbes modulaires,''
S\'eminaire Delange Pisot-Poitou, Th\'eorie des nombres {\bf 16}
(1974-1975), no. 1, talk no. 7, pp. 1-8.




\bibitem{dw} R. Donagi, K. Wendland, ``On orbifolds and free fermion
constructions,'' J. Geom. Phys. {\bf 59} (2009) 942-968,
{\tt arXiv:0809.0330}.

\bibitem{Sharpe:2019yag}
  E.~Sharpe,
  ``Categorical equivalence and the renormalization group,''
  {\tt arXiv:1903.02880}.



\bibitem{Niarchos:2018mvl}
  V.~Niarchos,
  ``Geometry of Higgs-branch superconformal primary bundles,''
  Phys.\ Rev.\ D {\bf 98} (2018) no.6,  065012,
  {\tt arXiv:1807.04296}.




\bibitem{diamonds} F. Diamond, J. Shurman, {\it A first course in modular
forms}, Graduate Texts in Math. 228,
Springer, New York, 2005.

\bibitem{s0} N. Seiberg, ``Modifying the sum over topological sectors
and constraints on supergravity,'' JHEP {\bf 1007} (2010) 070,
{\tt arXiv:1005.0002}.

\bibitem{fk} D. Freedman, B. K\"ors, ``K\"ahler anomalies in supergravity
and flux vacua,''
JHEP {\bf 0611} (2006) 067,
{\tt hep-th/0509217}.

\bibitem{efk} H. Elvang, D. Freedman, B. K\"ors, ``Anomaly cancellation
in supergravity with Fayet-Iliopoulos couplings,''
JHEP {\bf 0611} (2006) 068,
{\tt hep-th/0606012}.

\bibitem{yujipriv} Y. Tachikawa, private communication.

\bibitem{litt} D. Litt, ``Picard groups of moduli problems II,''
expository notes, available at
{\tt https://math.stanford.edu/\verb'~'dlitt/exposnotes/picardII.pdf}.

\bibitem{kconrad} K. Conrad, ``$SL(2,{\mathbb Z})$,''
available from \\
{\tt http://www.math.uconn.edu/\verb'~'kconrad/blurbs/grouptheory/SL(2,Z).pdf}.


\bibitem{fulton} W. Fulton, M. Olsson, ``The Picard group of
${\cal M}_{1,1}$,'' Alg. Num. Theory {\bf 4} (2010) 87-104.

\bibitem{mumford} D. Mumford, ``Picard groups of moduli problems,''
pp. 33-81 in {\it Arithmetical algebraic geometry (Purdue, December 1963)},
ed. O. F. G. Schilling, Harper \& Row, New York, 1965.

\bibitem{niles} A. Niles, ``The Picard groups of the stacks $Y_0(2)$
and $Y_0(3)$,''
Funct. Approx. Comment. Math. {\bf 55} (2016) 105-112,
{\tt arXiv:1504.07913}.

\bibitem{silverman} J. Silverman, {\it The arithmetic of elliptic
curves}, Springer-Verlag, New York, 1986.


\bibitem{katok} S. Katok, {\it Fuchsian groups}, University of Chicago
Press, Chicago, 1992.

\bibitem{silvermantate} J. Silverman, J. Tate, {\it Rational points on
elliptic curves}, Springer-Verlag, New York, 1992.





  

\end{thebibliography}
\end{document}